\documentclass[trackchanges,twocolumn]{aastex701}

\usepackage{enumitem}
\usepackage{amsmath}
\usepackage{booktabs}
\usepackage{graphicx}
\usepackage{subcaption}

\begin{document}

\title{3I/ATLAS: In Search of the Witnesses to Its Voyage}

\author[orcid=0000-0001-5797-252X]{X. P\'erez-Couto}
\affiliation{Universidade da Coruña (UDC), Department of Computer Science and Information Technologies, Campus de Elviña s/n, 15071, A Coruña, Galiza, Spain}
\affiliation{CIGUS CITIC, Centre for Information and Communications Technologies Research, Universidade da Coruña, Campus de Elviña s/n, 15071 A Coruña, Galiza, Spain}
\email[show]{xabier.perez.couto@udc.gal}  

\author[orcid=0000-0002-3150-8988]{S. Torres} 
\affiliation{Institute of Science and Technology Austria (ISTA), Am Campus 1, A-3400 Klosterneuburg, Austria}
\affiliation{Instituto de Astrofísica de Canarias, 38200 La Laguna, Tenerife, Spain}
\affiliation{Universidad de La Laguna (ULL), Astrophysics Department, 38206 La Laguna, Tenerife, Spain}
\email[show]{santiago.torres@iac.es}

\author[orcid=0000-0003-4936-9418]{E. Villaver} 
\affiliation{Instituto de Astrofísica de Canarias, 38200 La Laguna, Tenerife, Spain}
\affiliation{Universidad de La Laguna (ULL), Astrophysics Department, 38206 La Laguna, Tenerife, Spain}
\email{eva.villaver@iac.es}

\author[orcid=0000-0002-2086-3642]{A. J. Mustill}
\affiliation{Lund Observatory, Division of Astrophysics, Department of Physics, Lund University, Box 118, 22100, Lund, Sweden}
\email{alexander.mustill@fysik.lu.se}

\author[orcid=0000-0002-7711-5581]{M. Manteiga} 
\affiliation{Universidade da Coruña (UDC), Department of Nautical Sciences and Marine Engineering, Paseo de Ronda 51, 15011, A Coruña, Galiza, Spain}
\affiliation{CIGUS CITIC, Centre for Information and Communications Technologies Research, Universidade da Coruña, Campus de Elviña s/n, 15071 A Coruña, Galiza, Spain}
\email{minia.manteiga@udc.es}

\begin{abstract}

\noindent 3I/ATLAS is the third interstellar object discovered to date, following 1I/‘Oumuamua and 2I/Borisov. Its unusually high excess velocity and active cometary nature make it a key probe of the Galactic population of icy planetesimals. Understanding its origin requires tracing its past trajectory through the Galaxy and assessing the possible role of stellar encounters, both as a potential origin and a perturber to its orbit. We integrated the orbit of 3I/ATLAS backward in time for 10~Myr, together with a sample of \emph{Gaia}~DR3 stars with high-quality astrometry and radial velocities, to identify close passages within 2~pc. We identify 93 nominal encounters, 62 of which are significant at the $2\sigma$ level. However, none of these encounters produced any meaningful perturbation. The strongest perturber Gaia~DR3~6863591389529611264 at 0.30~pc and with a relative velocity of 35~km~s$^{-1}$, imparted only a velocity change of $|\Delta v| \simeq 5\times10^{-4}$~km~s$^{-1}$ to the orbit of 3I/ATLAS. Our results indicate that no stellar flybys within the past 10~Myr and 500~pc contained in \emph{Gaia}~DR3 can account for the present trajectory of 3I/ATLAS or be associated with its origin. We further show that 3I/ATLAS is kinematically consistent with a thin-disk population, despite its large peculiar velocity.

\end{abstract}


\section{Introduction} 

Kinematics is the key to identify interstellar objects (ISOs), since their trajectories provide the most direct evidence of an extrasolar origin. Unlike asteroids and comets formed within the Solar System, which remain confined to bound elliptical, or near-parabolic, orbits ($e\lesssim1$), ISOs arrive with velocities at infinity and thus follow hyperbolic trajectories ($e > 1$), with velocities determined by the Sun's motion through the Galaxy and the distribution of velocities of the ISO population. Such kinematic signatures provide a robust diagnostic, rendering ISOs clearly distinguishable from native Solar System populations.  

The discovery of the $\sim$100 m-sized 1I/'Oumuamua by the Pan-STARRS survey \citep{Meech2017} was long anticipated, as planet formation models predict the ejection of large amounts of material from the outer regions of forming planetary systems into interstellar space (e.g. \citealt{Raymond2018}). The efficiency of planetesimal ejection is highly sensitive to planetary architecture and dynamical history, yet dynamical models indicate that clearing processes are common across a wide range of system architectures \citep{Torres25}. As a result, interstellar space should be filled with planetesimals originating in protoplanetary disks, predominantly icy in composition given that the majority of ejected material is expected to form beyond the snow line of their parent systems.  

The discovery of 1I/'Oumuamua confirmed the existence of ISOs and highlighted their value as tracers of planet formation efficiency and the density of small bodies in the interstellar medium. Since then, two additional ISOs have been identified: 2I/Borisov \citep{deLeon2019} and, most recently, 3I/ATLAS \citep{Bolin2025, Seligman2025}. The latter exhibits an extremely hyperbolic heliocentric orbit with eccentricity $\sim$6.1 and an incoming velocity of $\sim$58 km s$^{-1}$, along with clear evidence of water-ice activity and CO$_2$ emission \citep{Xing2025,Lisse2025}, strongly suggesting an origin in the outer icy reservoir of another planetary system.  

Identifying the origin of ISOs is key to understanding planet formation efficiency, the distribution of volatiles and organics in the Galaxy, and the dynamical pathways by which planetary systems evolve. Proposed origins range from early ejection during planet formation \citep{Do2018,PZ_Torres2018}, to scattering in intermediate-age systems \citep{Brasser2013,Raymond2018,Torres2023}, to late-stage release by stellar perturbations or post-main-sequence mass loss \citep{JimenezTorres2011,Higuchi2015,Torres2019,Veras2011,Veras2012}. More exotic models have also been suggested, including tidal fragmentation of comets \citep{ZhangLin2020}, formation in molecular cloud cores \citep{Levine2021}, and tidal disruption of asteroids around white dwarfs \citep{Rafikov2018}.  

The unusually high excess velocity of 3I/ATLAS has already motivated attempts to constrain its origin using age--velocity relations and Galactic dynamics \citep{Hopkins2025ATLAS,Taylor2025,delafuenteMarcos2025,kakharov2025, guo2025}. If the chemical abundances and Galactic dynamics of a sample of ISOs are both sufficiently constrained, it would allow tests of correlations between kinematics and compositions expected from Galactic chemodynamics models, and hence allow us to probe the compositions of planetary building blocks across space and time in the Galaxy \citep{Hopkins2023,Hopkins2025}. In a very fortuitous case, a reconstruction of the ISO's past Galactic orbit, coupled with those of nearby stars, could reveal the original parent star of the ISO
. Moreover, encounters with other stars can gravitationally perturb ISO trajectories \citep[e.g.,][]{Heisler1986,Rickman2014,Feng2015}, changing their Galactic orbit. Tracing ISO orbits back through the Galaxy remains challenging, however, as small uncertainties in orbital elements and stellar astrometry grow rapidly over time \citep{Zhang2018}, even with \emph{Gaia} \citep{2016A&A...595A...1G} precision. {In addition to astrometric uncertainties, the reconstruction of ISO trajectories is fundamentally limited by uncertainties in the Galactic potential itself, including non-axisymmetric and time-dependent structures such as spiral arms and giant molecular clouds, which render deterministic orbit tracing over tens to hundreds of Myr unreliable in principle.

In this work, taking these constraints and limitations into account, we focus on maximizing the reliability of our results by reconstructing the recent history of stellar encounters experienced by 3I/ATLAS over the past 10 Myr, prior to its entry into the Solar System. We use \emph{Gaia} DR3 astrometry to identify and model encounters with known stars currently located within 500 pc of the Sun.} Section~\ref{sec:methods} describes the methodology. Section~\ref{sec:results} presents the closest encounters their dynamical effects on the orbit of 3I/ATLAS, and in Section \ref{sec:discussion} we discuss the results. Section~\ref{sec:conclusions} summarizes our conclusions.  

\section{Methods}
\label{sec:methods}

To trace the past trajectory of 3I/ATLAS, we adopted the position and velocity computed by \citet{delafuenteMarcos2025} at an epoch prior to its entry into the Solar System ($\approx 3 \times 10^{4}$ yr ago), namely: $\alpha = 295^{\circ}.043^{+0.003}_{-0.004}$, $\delta = -19^{\circ}.0704^{+0.0006}_{-0.0005}$, a heliocentric distance of $1.7819 \pm 0.0003$ pc, and Cartesian Galactic heliocentric velocities $(U, V, W) = (-51.233 \pm 0.006, -19.456 \pm 0.004, +18.930 ^{+0.005}_{-0.006})$ km s$^{-1}$.

The potential stellar encounters are identified by selecting \emph{Gaia}~DR3 sources with high-quality astrometry within 500 pc of the Sun. Only sources satisfying:
\begin{itemize}[noitemsep,topsep=1pt,parsep=0pt,partopsep=1pt]
    \item \texttt{parallax} $> 2$ mas
    \item \texttt{parallax\_over\_error > 10}
    \item \texttt{ruwe < 1.4}
    \item \texttt{visibility\_periods\_used} $\geq 8$
\end{itemize}
were retained from the \emph{Gaia} Archive query, yielding a total of 13,896,270 stars. Since high-quality radial velocities (RVs) are essential for our kinematic analysis, we supplemented the RVs from the \emph{Gaia} Radial Velocity Spectrometer \citep[RVS;][]{Katz2023} with additional measurements from APOGEE-2 DR17 \citep{Abdurrouf2022}, LAMOST DR7 \citep{Luo2019}, GALAH DR3 \citep{Buder2021}, RAVE DR6 \citep{Steinmetz2020}, and the \emph{Gaia}-ESO DR5 survey \citep{Hourihane2023}. {Most of these surveys already provide a \emph{Gaia} source identifier associated with each RV measurement. For LAMOST and \emph{Gaia}-ESO, however, we performed a positional cross-match using equatorial coordinates with a search radius of $0.5''$.}

If multiple RV measurements are available for a source, we adopt the one with the lowest associated uncertainty, provided it is within $10$ km s$^{-1}$ of the median RV for that source. Otherwise, the \emph{Gaia} RVS is provided. Our final sample therefore consists of 3,608,022  sources with 6D initial conditions\footnote{Since our analysis is restricted to nearby stars ($< 500$ pc) with small fractional parallax errors ($\sigma_{\varpi}/\varpi<0.1$), we can safely estimate distances $d$ using the inverse of the parallax ($1/\varpi$)}. We corrected the \emph{Gaia} parallaxes using \citet{Lindegren2021} and the proper motions following  \citet{Cantat-Gaudin2021}, and increased the parallax uncertainties following \citet{MaizApellaniz2022}.

To perform the orbital traceback, we used \texttt{Gala}, a Python package for galactic dynamics \citep{Price2017}. For the Galactic potential, we adopted \texttt{MilkyWayPotential2022}, which includes a spherical nucleus and bulge, a sum of Miyamoto-Nagai disks, and a spherical NFW halo, fit to the \cite{Eilers2019} rotation curve. To transform the heliocentric reference frame into the galactocentric one, we adopted the solar distance $R_0$ to the galactic center as 8.122 kpc \citep{GRAVITY2018}, the Solar velocity relative to the Local Standard of Rest (LSR) as $(U_\odot,V_\odot,W_\odot) = (11.1, 12.24, 7.25)$ km s$^{-1}$ \citep{Schonrich2010}, an LSR velocity of $V_C(R_0) = 229$ km s$^{-1}$ \citep{Eilers2019}, and a Solar height above the Galactic plane of $Z_\odot = 20.8$ pc \citep{Bennett2019}.

We first integrated the orbits of both 3I/ATLAS and the \emph{Gaia} DR3 sources backward in time for $100$ Myr, {with $10^4$ time steps of $\Delta t = 0.01 $ Myr, and using the \texttt{LeapFrog} integrator.}
Because we found excessively large errors beyond $-10$ Myr, we repeat the integration up to $t = -10$ Myr and with a finer time step ($\Delta t = 1000$ yr). We also performed a longer 12\,Gyr integration of the orbit of 3I/ATLAS to ensure we had captured its maximum $Z-$excursions.

To overcome memory and computational limitations, we split the bulk sample into onion-like layers with outer boundaries at 100, 150, 200, 250, 300, 350, 400, 450, and 500 pc and progressively performed our calculations on each layer. A potential close encounter is defined as the time step $t$ at which the Euclidean distance between 3I/ATLAS and a given star ($d_{\rm rel}$) falls below the adopted \emph{critical radius}, $r = 2$~pc, corresponding to the maximum distance at which a passing star can significantly perturb a comet \citep[e.g.,][]{Torres2019,PortegiesZwart2021}. This is also roughly the same size as the Hill radius where the star's gravity dominates over the Galactic potential in a corotating frame. We identified 93 nominal stellar encounters (after discarding 14 sources with $|RV| > 500$ km s$^{-1}$). 

To account for uncertainties in the initial conditions, arising primarily from the RVs, we re-integrated the orbits of these candidate stars, drawing $10^3$ Monte Carlo (MC) realizations of their orbits from Gaussian distributions of their uncertainties, including covariances between the astrometric parameters. For 3I/ATLAS, where some solutions show asymmetric errors, we used the largest one for the sampling. Finally, we defined high-confidence close encounters as those for which at least $95\%$ of the sampled orbits approached within the critical radius.

\section{Encounter history of 3I/ATLAS} 
\label{sec:results}

\begin{figure}[t]
  \centering
  \begin{subfigure}[t]{0.76\columnwidth}
    \centering
    \includegraphics[width=\linewidth]{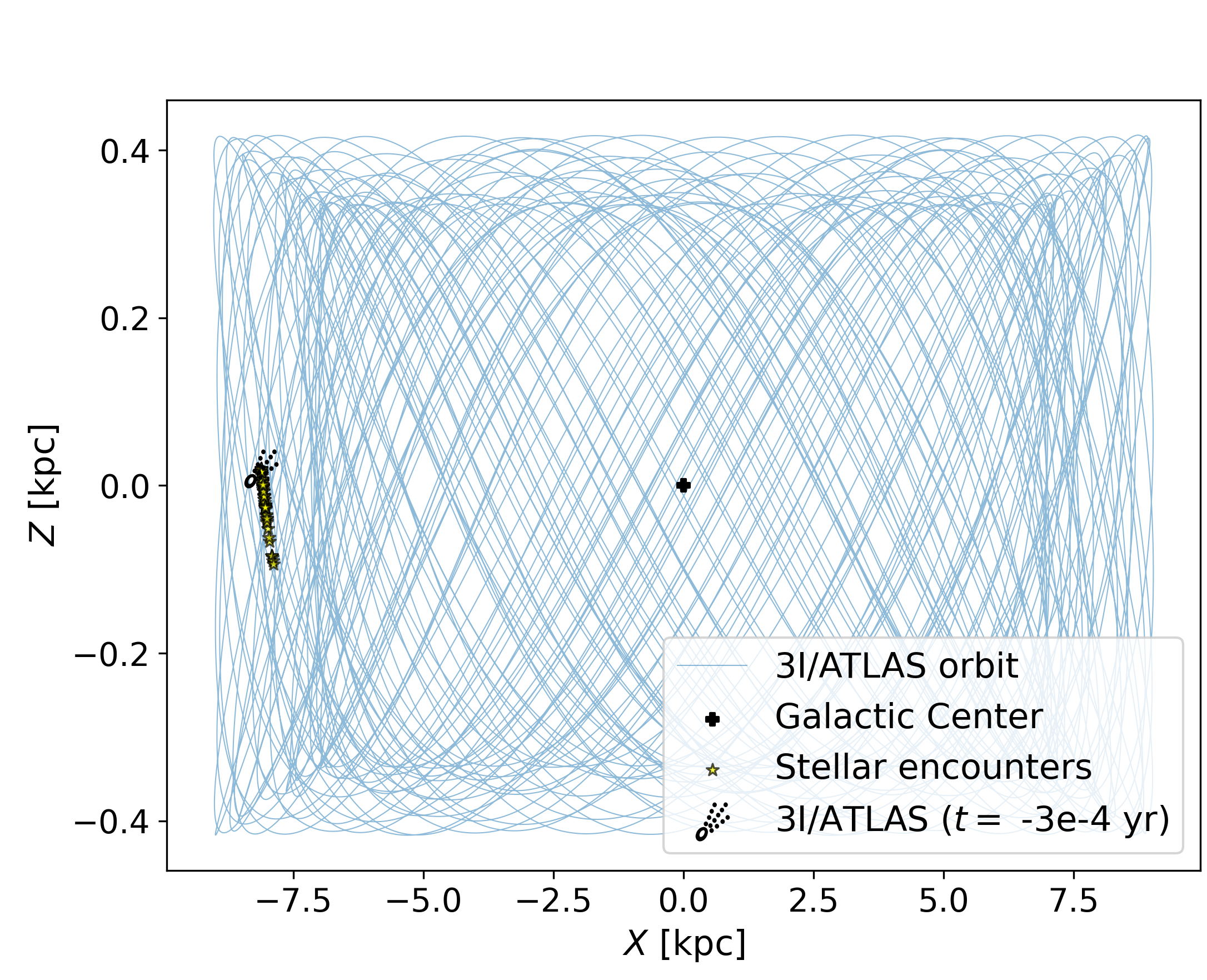}
  \end{subfigure}\hfill
  \begin{subfigure}[t]{0.229\columnwidth}
    \centering
    \includegraphics[width=\linewidth]{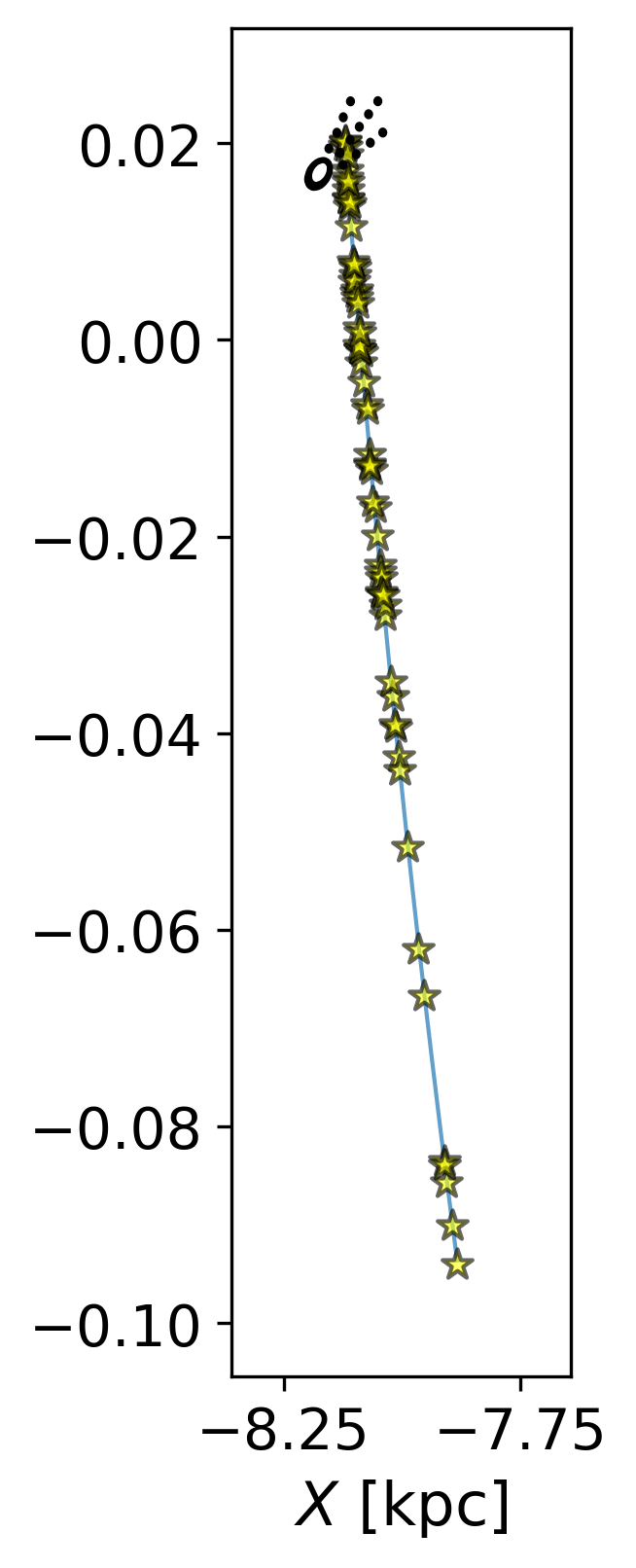}
  \end{subfigure}

  \caption{Galactic trajectory of 3I/ATLAS integrated 12 Gyr (\textit{left panel}) and $\sim 4$ Myr (\textit{right panel}) backward in time, projected onto the Galactic XZ plane. Stars mark the 62 high-confidence encounters identified in this work. The error bars for the stars are not visible, as they are smaller than the plotted symbols.}
  \label{fig:close_encounters_XZ}
\end{figure}

After propagating the orbit of 3I/ATLAS for 12\,Gyr the object reached a maximum Galactic altitude of $|Z| \approx 420$ pc, placing it near the boundary of the thin disk \citep{Vieira2023, Tian2024}. Figure \ref{fig:close_encounters_XZ} shows the close encounters with respect to the 3I/ATLAS' orbit projected in the XZ Galactic plane. Our Galactic trajectory is in strong agreement with that computed by \citet{kakharov2025}.

We identified 62 high-confidence stellar encounters of 3I/ATLAS in the first 10 Myr (Table \ref{tab:encounters}).
To estimate the dynamical influence of these stellar flybys, we adopt the  \emph{Classical Impulse Approximation} (CIA; \citealt{Rickman1976,Galactic_dynamics08}).  Although based on a simple assumption, the CIA provides a  useful first-order diagnostic for assessing whether 3I/ATLAS  experienced an encounter strong enough to significantly alter its  trajectory, without the need of performing a full $N$-body integration. This approach assumes rectilinear stellar motion and treats the encounter as an instantaneous perturbation, which is valid given that the duration of the encounter is short compared to the comet’s dynamical timescale.  In this framework, a passing star of mass $M_\star$ and relative  velocity $v_{\rm rel}$ imparts an instantaneous change in the velocity of the comet at closest approach.  For an encounter with impact parameter $d_{\rm rel}$,  the instantaneous velocity kick imparted during the flyby, $\Delta v$, can be expressed as, $|\Delta v| \;=\; \frac{2GM_\star}{d_{\rm rel}\,v_{\rm rel}}$, where $G$ is the gravitational constant. It is convenient to express the encounter strength in a dimensionless form by normalizing the impulse to the relative velocity, which yields the effective deflection angle of the hyperbolic trajectory, $\theta \;\equiv\; \frac{|\Delta v|}{v_{\rm rel}}$ \citep{Galactic_dynamics08}. Here, $\theta$ denotes the angular deflection of 3I/ATLAS’ velocity vector caused by the flyby. For typical relative velocities of $v_{\rm rel} \sim 20$--$60$ km s$^{-1}$, the value of $\theta$ quantifies the dynamical strength of the encounter, ranging from  very weak ($10^{-6} \leq \theta < 10^{-5}$), through moderate ($10^{-4} \leq \theta < 10^{-3}$), to very strong
$\theta \geq 10^{-2}$ (corresponding to au-scale flybys). 
These values correspond to impact parameters of $\sim 0.5$~pc for very weak encounters, and$\sim 10^3$~au for strong ones.  Thus, only encounters within $\lesssim 10^3$--$10^4$~au and/or with  very low relative velocities can produce dynamically significant  deflections \citep{Torres2019}. At parsec-scale distances, even solar-mass stars yield  $\theta \sim 10^{-6}$, corresponding to sub-m~s$^{-1}$ perturbations which are negligible for 3I/ATLAS’ Galactic trajectory. 

\begin{figure}
    \centering
    \includegraphics[width=0.5\textwidth]{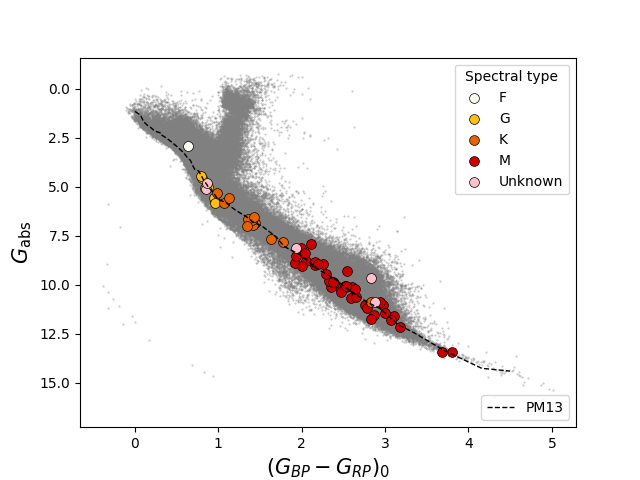}
    \caption{\textit{Position on the \emph{Gaia}} CMD diagram of the stars that experienced a close encounter with 3I/ATLAS, with spectral types indicated in the legend that comes from \citet{Creevey2023}. The background shows all the sources from our \emph{Gaia} DR3 500 pc sample with available RVs, after applying the same magnitude and parallax error cuts as for the encounter stars. The dashed line marks the MS-locus defined in \citep[][PM13 in the legend]{PecautMamajek2013}.}
    \label{fig:close_encounters_HR}
\end{figure}

We computed $|\Delta v|$ and $\theta$, deriving stellar masses via cubic-spline interpolation of the de-reddened color $(G_{BP} - G_{RP})_0$ using the updated \citet{PecautMamajek2013} tables\footnote{\url{https://www.pas.rochester.edu/~emamajek/EEM_dwarf_UBVIJHK_colors_Teff.txt}}. To deal with the extinction and reddening we used the package \texttt{dustmaps} \citep{Green2018} and the \citet{Leike2020} 3D dust map to obtain the color excess $E(B-V)$ for each source. Subsequently, we used the \texttt{dust-extinction} \citep{Gordon2024} library and the \citet{Gordon2023} extinction law \citep[based on the previous works of][]{Gordon2009, Fitzpatrick2019, Gordon2021, Decleir2022} to obtain the final values.
As shown in Figure \ref{fig:close_encounters_HR}, all the stars with close encounters lie on the main sequence (MS), and most of them are already identified as so in the Extended Stellar Parametrizer for Hot Stars \citep[ESP-HS,][]{Creevey2023}.

In Table 1, we report the deflection angles produced by close stellar encounters within 500 pc over the past 10 Myr on the trajectory of 3I/ATLAS. We find deflection angles on the order of $10^{-5}$--$10^{-6}$ for most of the stars, showing that none of these encounters had a significant impact on 3I/ATLAS’s trajectory.

The strongest encounter in our sample was with the star  \emph{Gaia}~DR3~6863591389529611264, which passed within 0.30~pc of 3I/ATLAS at a relative velocity of 35~km~s$^{-1}$, about 72,000~yr ago. This flyby induced a velocity change of only  $\Delta v \simeq 5\times10^{-4}$~km~s$^{-1}$, corresponding to a deflection angle of $\theta \simeq 1.6\times10^{-5}$. Equivalently, this would result in a position deviation of only $\sim 0.05$ pc after 100 Myr. Such a small perturbation indicates that 3I/ATLAS has not been significantly influenced by encounters with stars with known kinematics over the past 4.27~Myr (the epoch of the earliest encounter identified in this work). 

\begin{figure}
    \centering
    \includegraphics[width=0.5\textwidth]{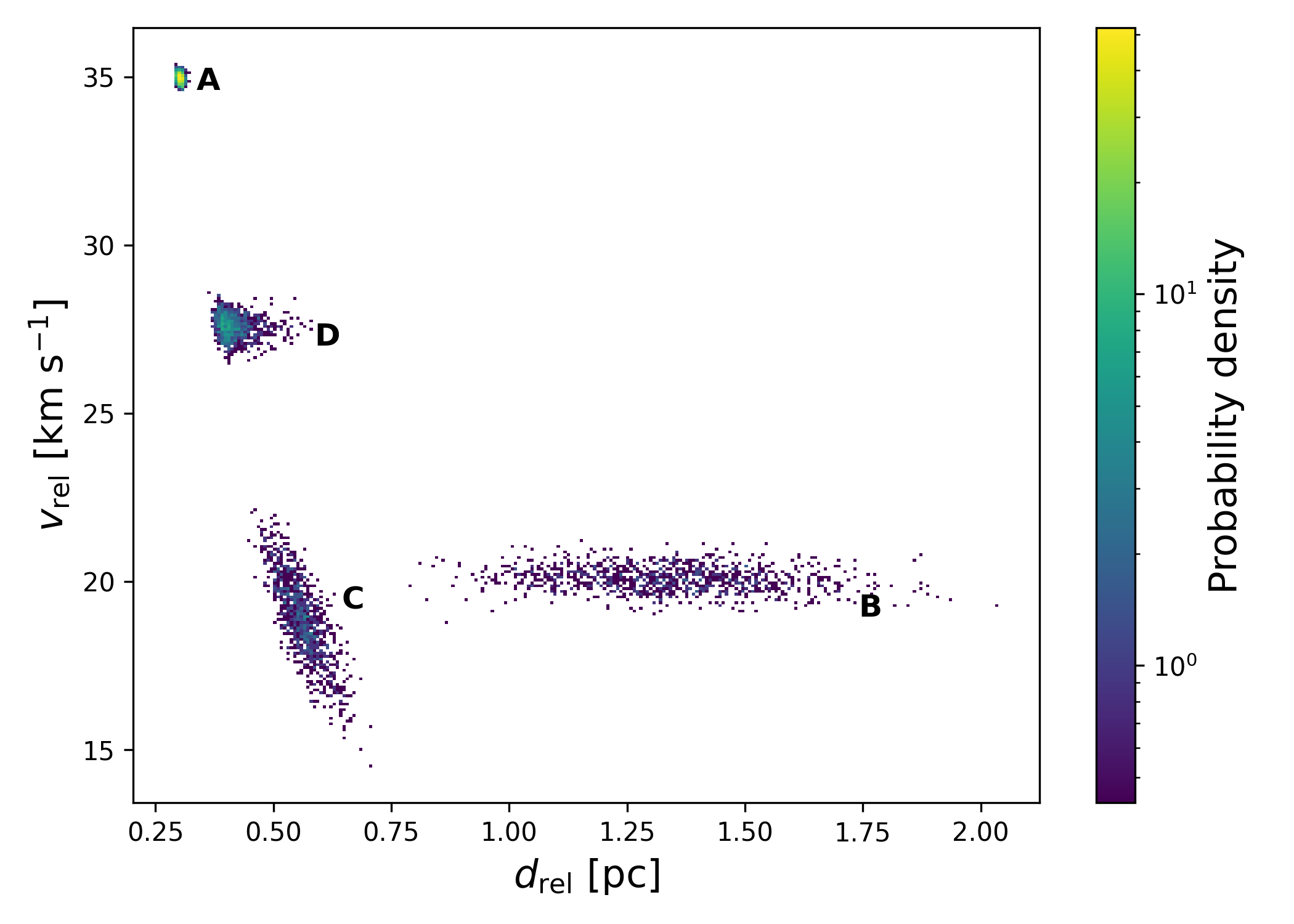}
    \caption{Relative velocity $v_{\rm rel}$ as a function of relative distance $d_{\rm rel}$ for the four strongest encounters obtained from $10^3$ Monte Carlo orbits. {The color scale shows the probability density, estimated from a normalized 2D histogram, in units of pc$^{-1}$ (km s$^{-1}$)$^{-1}$}.}
    \label{fig:vrel_drel}
\end{figure}

For clarity, {we present in Figure \ref{fig:vrel_drel} a finite approximation of the probability density-distribution of $v_{\rm rel}$ and $d_{\rm rel}$ between 3I/ATLAS and the four strongest encounters (those with the greater $\theta$). {We computed them as normalized 2D histograms (using $250 \times 250$ bins) of the $10^3$ MC realizations for each encounter, where the normalization ensures that the integral over the full parameter space equals unity. The encounters are}, in descending order of $\theta$,} \emph{Gaia} DR3 6863591389529611264 (A, $M_{A} \approx 0.70 M_{\odot}$), \emph{Gaia} DR3 6731311275891314304 (B, $M_{B} \approx 0.98 M_{\odot}$), \emph{Gaia} DR3 1197546390909694720 (C, $M_{C} \approx 0.52 M_{\odot}$), and \emph{Gaia} DR3 4591398521365845376 (D, $M_{D} \approx 0.20 M_{\odot}$). As Figure \ref{fig:vrel_drel} illustrates, although they involve very small relative distances, their relative velocities are always greater than $15-20$ km\,s$^{-1}$.

\begin{figure}
    \centering
    \includegraphics[width=0.5\textwidth]{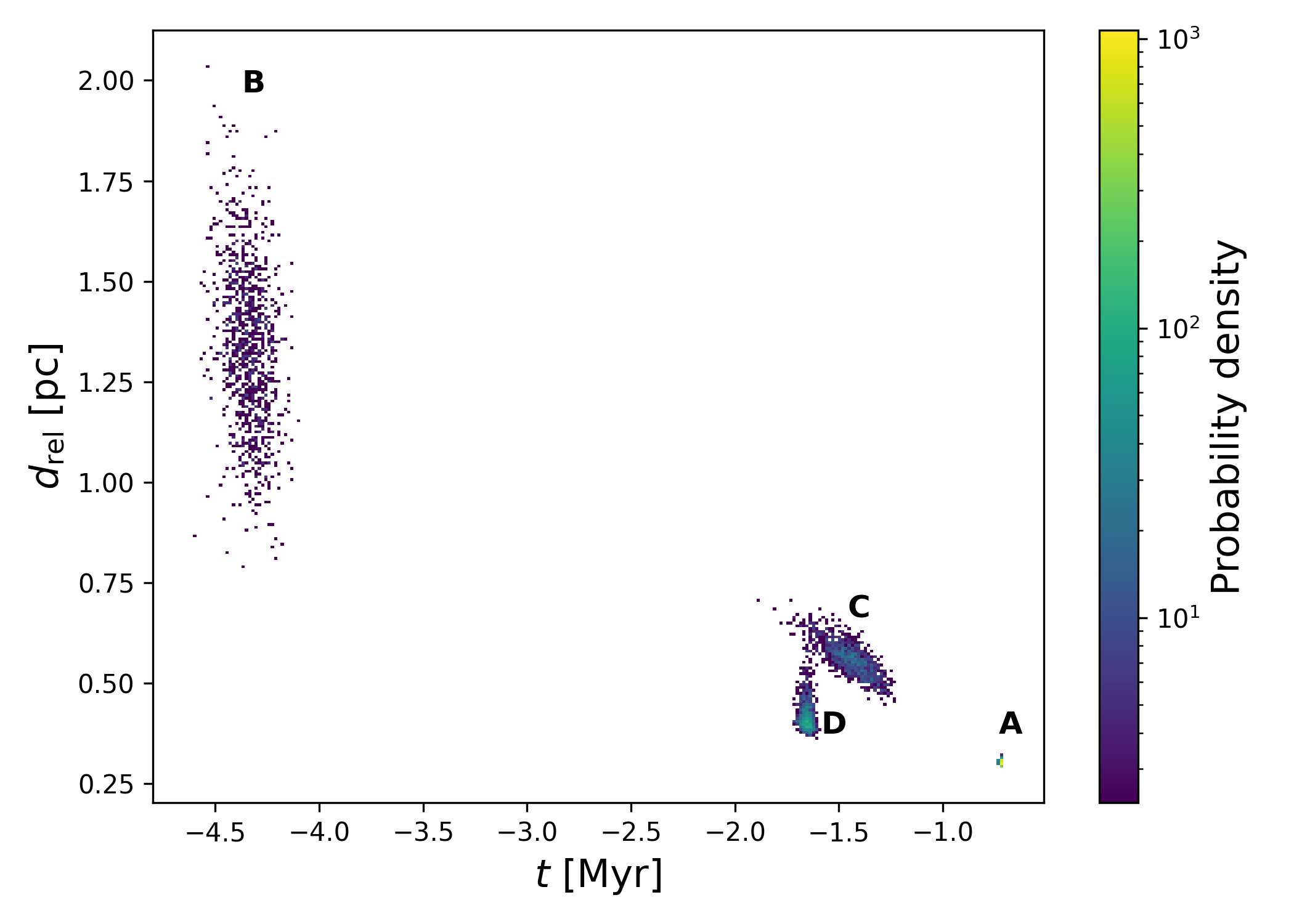}
    \caption{Relative distance $d_{\rm rel}$ as a function of time $t$ for the four strongest encounters obtained from $10^3$ Monte Carlo orbits. {The color scale shows the probability density, estimated from a normalized 2D histogram, in units of Myr$^{-1}$ pc$^{-1}$}.}
    \label{fig:drel_t}
\end{figure}

In addition to this, Figure \ref{fig:drel_t} shows the probability density distribution of $d_{\rm rel}$ as a function of encounter time $t$. The dispersion in relative distance increases with the integration time, especially for $t <-1$ Myr (as seen for encounter B).

\subsection{Completeness of the Encounter Sample}

To assess the number of stellar encounters that may have been missed—primarily due to the lack of precise RV measurements—we followed the procedure described in \citet{grageramas2025}.

As a first approximation, the total number of stellar encounters $N$ expected over a time span $t$, within a cylindrical volume of cross section $\pi r^2$, can be estimated from the stellar density in the solar neighborhood $\rho_{\rm ST}$ (which, as is well known, depends strongly on the spectral type), and their characteristic encounter velocity $v_{\rm ST}$. The latter is given by $v_{\rm ST} = \sqrt{v_{\rm 3I/ATLAS, ST} ^2 + \sigma _{\rm ST}^2}$, where $v_{\rm 3I/ATLAS, ST}$ is the mean relative velocity of the stars of a given spectral type and 3I/ATLAS, and $\sigma _{\rm ST}$ is the velocity dispersion of that spectral type.

Therefore, we estimate $N$ within $r=2$ pc and a time span of $t=4.27$ Myr (the time interval between the first and last close encounter identified in our work), as 

\begin{equation}\label{eq:cylindrical_approx}
N = \sum_{\rm ST} \pi r^2 \, t \, v_{\rm ST}\,  \rho_{\rm ST}. 
\end{equation}

{To obtain $v_{\rm 3I/ATLAS, ST}$ we transformed the values of $v_{\rm Sun, ST}$ (that is, the mean relative velocity of stars with spectral type ST with respect to the Sun) computed in \citet{Rickman2008} to the 3I/ATLAS's frame by subtracting its 3D velocity. For $\rho_{\rm ST}$ and $\sigma _{\rm ST}$ we used the values provided by \citet{Torres2019}. As a result, we obtained $N=246$ expected encounters,} of which we are reporting approximately $\sim 25 \%$. This fraction is pretty close to the number of RVs available in our sample ($\sim 31\%$) within the heliocentric distance of the most distant encounter ($\sim 245$ pc). 

Although we recognize that the lack of RVs is a limitation for the number of integrated orbits, it is also true that it is highly unlikely that any of the remaining 75\% of encounters would affect the orbit of 3I/ATLAS in this time window. Indeed, if we assume $r = d_{\rm rel}$ we can combine the CIA, the effective deflection angle of the hyperbolic trajectory equations, and Equation \eqref{eq:cylindrical_approx} to obtain the expected number of encounters for a given deflection angle $\theta$. It is therefore easy to see that the expected number of strong encounters ($\theta > 10^{-2}$ rad) is nearly zero ($N = 10^{-5}$) for the $t=4.27$ Myr time span. {Furthermore, this equals to an encounter rate of $\sim 10^{-14}$~yr$^{-1}$, which is in consonance with the results of \citet{Forbes2025}.}

\section{DISCUSSION}
\label{sec:discussion}

Understanding the origin of 3I/ATLAS is relevant as it could provide rare empirical constraints on how interstellar objects are produced and evolve, testing competing formation scenarios—from tidal disruptions around white dwarfs to exo-Oort cloud ejections—and, through its kinematics, it directly informs our picture of how planetary material {is ejected, transported, and redistributed across the Galaxy.}

Throughout our kinematic study of 3I/ATLAS within the Galactic potential, we investigated whether it could have been ejected by a nearby star or significantly perturbed by a stellar encounter capable of imparting its present peculiar velocity. By integrating the orbit of 3I/ATLAS 10 Myr backward with \emph{Gaia} DR3 stars, we identified 93 nominal encounters (62 at the $2\sigma$ level), yet none were capable of producing significant perturbations (Fig. \ref{tab:encounters}). Even the closest \emph{Gaia} DR3 6769021226194779136 (0.27 pc, 0.19~M$_\odot$) and strongest \emph{Gaia} DR3 6863591389529611264 (0.30 pc, 0.7~M$_\odot$) encounter occurred at high relative velocities, leaving 3I/ATLAS’s origin unassociated with any specific stellar system. We conclude that 3I/ATLAS has not experienced any stellar flybys within the past 4 Myr, among the stars contained in \emph{Gaia} DR3, that could account for its present trajectory nor its origin. There is a caveat this analysis provided possible incompleteness, that we estimated from \emph{Gaia}’s coverage of the local 245 pc volume and found it to be only $\sim$25\% complete. However, for the remaining 75\%, we present a first approximation analysis based on statistics within 100 pc of the Sun. This analysis shows that a strong encounter ($\theta > 10^{-2}$) was unlikely to occur in any case. This is expected on our understanding of kinematic heating of the Galactic disk, which is caused not by star--star encounters but by their interaction with larger agglomerations of matter such as GMCs, satellite galaxes and spiral arms \citep[e.g.,][]{Lacey1984,TothOstriker1992,AumerBinney2009}.

We further examine the orbit of 3I/ATLAS by analyzing the extent of its vertical motion in the Galactic disk, and conclude that it most likely originates from the thin disk. Its vertical excursion reaches only  $|Z|\sim0.42$~kpc (Fig. \ref{fig:close_encounters_XZ}), consistent with the thin-disk scale height and far smaller than the kiloparsec-scale typical of thick-disk stars \citep[e.g.,][]{Gilmore1983,Bensby2003,Recio-Blanco2014}.

As noted by \citet{delafuenteMarcos2025}, the space motions $(U,V,W)=(-51.2,-19.5,+18.9)\ \mathrm{km\,s^{-1}}$ correspond to a Toomre velocity of $T\approx58\ \mathrm{km\,s^{-1}}$, well below the canonical thin/thick-disk division at $T\sim70$--$100\ \mathrm{km\,s^{-1}}$ \citep[e.g.,][]{Bensby2003,Reddy2006,Recio-Blanco2014}. In addition, applying the \citealt{Bensby2014} criteria yields an odds ratio of TD/D = 0.04 for membership of the thick versus thin disk, implying that 3I/ATLAS is about 20 times more likely to belong to the thin disc than to the thick disc based purely on kinematics (Appendix. \ref{kinematic}). While this strongly favors thin-disk kinematics, \citet{Bensby2014} note that purely kinematic criteria can misclassify a minority of stars, some objects with ${\rm TD/D}<0.1$ exhibit chemical abundance patterns characteristic of the thick disk (e.g.\ $\alpha$-enhancement, low [Fe/H]). In the absence of chemical information for 3I/ATLAS, its classification  remains probabilistic. Nevertheless, when combined with its  vertical component ($|Z|\sim0.42$ kpc) and its location in the Toomre  diagram ($T\approx55$ km s$^{-1}$), the Bensby analysis consistently supports a thin-disk origin for 3I/ATLAS.

\citet{Taylor2025} used the age--velocity dispersion relation to infer a median kinematic age of $\sim7$~Gyr for 3I/ATLAS, with a 68\% confidence interval of 3--11~Gyr, which is consistent with both a thick disk and thin disk origin. Our classification relies on the present-day phase-space location of 3I/ATLAS relative to Galactic populations, whereas \citet{Taylor2025} infers a statistical age. Taken together, these results indicate that while 3I/ATLAS follows a thin-disk orbit in the solar neighborhood, it may nonetheless be an old object, consistent with ejection from a long-lived primordial planetesimal disk in an early-formed system. 

The recent study \citet{guo2025} examined the past stellar encounters of 3I/ATLAS. Although they apply a similar methodology, their list of 25 encounters overlaps with ours in only 13 cases. To address this discrepancy, we re-integrated the orbits of the 20 single sources in common with our initial sample. We find that, while the relative velocities and time of encounters agree with their results, the relative distances exceed $2$ pc for 10 of these stars (and the remaining 2 shows a RUWE $> 2$). To understand and quantify the source of the differing results, in Appendix \ref{comparison_guo} we examine three key methodological differences between this study and that of \citet{guo2025}: i) the initial conditions; ii) the \emph{Gaia} systematics and iii) the adopted galactic potential. We find good agreement with \citet{guo2025} for encounters within the past 1~Myr, whereas at longer look-back times our results diverge, with encounter distances drifting as the integration time increases. This indicates that the robust encounters primarily reflect the astrometric precision of \emph{Gaia} on short timescales, while the discrepancies at earlier epochs arise from the sensitivity of the results to the adopted Galactic potential and mass model.

\section{Summary and Conclusions} 
\label{sec:conclusions}

The emerging picture from recent studies is that interstellar objects are a heterogeneous population that provide rare observational constraints on planetary system evolution across the Galaxy. For 3I/ATLAS, we find that the 62 close stellar encounters identified in \emph{Gaia} DR3 within 500~pc over the past 10~Myr are too fast and at too large distances to have shaped its current orbit. None of them produced deflections larger than $10^{-5}$ radians, leaving 3I/ATLAS's trajectory essentially unaffected. We estimated the possible incompleteness of our analysis from \emph{Gaia}’s coverage of the local 245 pc volume and found it to be only $\sim$25\% complete. We also found supporting dynamical arguments that favor a thin-disk origin of 3I/ATLAS. Together, all data indicate that while 3I/ATLAS follows a thin-disk orbit in the solar neighborhood, it may nonetheless be an old object, consistent with ejection from a primordial planetesimal disk in an early-formed system, or from an exo-Oort cloud, and is most likely associated with the transition region between the thin and thick disk, although its origin remains undisclosed.

Our study shows that the encounter history of individual ISOs remains, despite the exquisite astrometric precision of \emph{Gaia,} incomplete and subject to systematic errors in the astrometry and Galactic potential even within a few Myr. Nevertheless, chemodynamic trends may still be revealed once a large sample of ISOs has been observed.

\begin{acknowledgments}

{We thank the anonymous referee for a careful reading of the manuscript and for constructive comments that improved the paper.}
XPC and ST thank J.L. Gragera-Más and Ylva Götberg, for their valuable feedback and comments. XPC acknowledges financial support from the Spanish National Programme for the Promotion of Talent and its Employability grant PRE2022-104959 cofunded by the European Social Fund. ST acknowledges the funding from the European Union’s Horizon 2020 research and innovation program under the Marie Sk\l{}odowska-Curie grant agreement No 101034413.
EV acknowledges support from the \textit{DISCOBOLO} project funded by the Spanish Ministerio de Ciencia, Innovaci\'on y Universidades under grant PID2021-127289NB-I00. AJM acknowledges support from the Swedish National Space Agency (Career grant 2023-00146). XPC and MM acknowledge support from the Spanish Ministerio de Ciencia, Innovaci\`on y Universidades under grants PID2021122842OB-C22 and PID2024-157964OB-C22; from the Xunta de Galicia and the European Union (FEDER Galicia 2021-2027 Program) Ref. ED431B 2024/21, ED431B 2024/02, and CITIC ED431G 2023/01. This work has made use of data from the European Space Agency (ESA) Gaia mission and processed by the Gaia Data Processing and Analysis Consortium (DPAC). Funding for the DPAC has been provided by national institutions, in particular, the institutions participating in the Gaia Multilateral Agreement.

\end{acknowledgments}




\software{astropy \citep{2013A&A...558A..33A,2018AJ....156..123A,2022ApJ...935..167A}, Matplotlib \citep{Hunter:2007}, NumPy \citep{harris2020array}, Pandas \citep{reback2020pandas}}

\appendix

\section{3I/ATLAS Close Encounters}
\label{table_encounters}

Table \ref{tab:encounters} lists the high-confidence stellar encounters identified in this study.

\begin{table}[!ht]
    \centering
    \caption{\label{tab:encounters} Identified stellar encounters within $2$ pc of 3I/ATLAS at the $2 \sigma$ level. For each encounter, we list clouds involved,  time of closest approach $t_{\rm closest}$,  median relative distance $d_{\rm rel}$ and velocity $v_{\rm rel}$ with $1\sigma$ uncertainties. The final columns gives the median encounter strength $|\Delta v|$, and deflection angle $\theta$.}
    \scriptsize

    \begin{tabular}{lccccccc}
            \toprule
            \emph{Gaia} DR3 ID & ST & $M_{\ast}$ & $t_{\rm closest} \pm \sigma_{t_{\rm closest}}$ & $d_{\rm rel} \pm \sigma_{d_{\rm rel}}$ & $v_{\rm rel} \pm \sigma_{v_{\rm rel}}$ & $|\Delta v|$ & $\theta$  \\ 
             & &[$M_\odot$] & [Myr] & [pc] & [km s$^{-1}$] & [km s$^{-1}$] & [rad] \\ \midrule
              6769021226194779136 & M & 0.19 & $-0.947\pm0.004$ & $0.276\pm0.016$ & $72.11\pm0.17$ & 8.37e-05 & 1.16e-06 \\
        6863591389529611264 & K & 0.70 & $-0.720\pm0.003$ & $0.303\pm0.005$ & $34.97\pm0.13$ & 5.71e-04 & 1.63e-05 \\
        6779821003058453120 & M & 0.18 & $-0.331\pm0.006$ & $0.397\pm0.008$ & $28.33\pm0.47$ & 1.38e-04 & 4.87e-06 \\
        5944464849163504128 & K & 0.83 & $-1.630\pm0.010$ & $0.407\pm0.136$ & $67.54\pm0.32$ & 2.60e-04 & 3.84e-06 \\
        1197546390909694720 & M & 0.52 & $-1.654\pm0.020$ & $0.408\pm0.033$ & $27.59\pm0.33$ & 4.00e-04 & 1.45e-05 \\
        4484666797357868288 & G & 1.05 & $-3.910\pm0.046$ & $0.504\pm0.407$ & $47.10\pm0.51$ & 3.79e-04 & 8.05e-06 \\
        6855915149098312064 & M & 0.25 & $-0.583\pm0.056$ & $0.550\pm0.052$ & $105.18\pm9.58$ & 3.67e-05 & 3.49e-07 \\
        1197546562708387584 & M & 0.40 & $-1.688\pm0.057$ & $0.555\pm0.076$ & $27.03\pm0.90$ & 2.29e-04 & 8.49e-06 \\
        4591398521365845376 & M & 0.20 & $-1.440\pm0.095$ & $0.561\pm0.041$ & $18.91\pm1.22$ & 1.63e-04 & 8.64e-06 \\
        6482924967749137152 & F & 1.25 & $-2.686\pm0.019$ & $0.582\pm0.353$ & $49.04\pm0.21$ & 3.77e-04 & 7.68e-06 \\
        6773261320990713216 & M & 0.32 & $-1.293\pm0.127$ & $0.583\pm0.073$ & $68.00\pm6.34$ & 6.96e-05 & 1.02e-06 \\
        6620396322451894656 & -- & 0.55 & $-3.983\pm0.274$ & $0.591\pm0.404$ & $31.82\pm2.10$ & 2.50e-04 & 7.86e-06 \\
        2386898972054841088 & M & 0.54 & $-0.534\pm0.003$ & $0.598\pm0.004$ & $38.60\pm0.19$ & 2.00e-04 & 5.18e-06 \\
        692654877680490880 & M & 0.43 & $-2.154\pm0.965$ & $0.670\pm6.487$ & $60.34\pm17.12$ & 9.10e-05 & 1.51e-06 \\
        6640455159755767424 & M & 0.38 & $-2.368\pm0.625$ & $0.712\pm0.526$ & $43.64\pm9.32$ & 1.05e-04 & 2.41e-06 \\
        6745564794882995712 & M & 0.36 & $-1.178\pm0.063$ & $0.712\pm0.057$ & $90.76\pm4.75$ & 4.79e-05 & 5.27e-07 \\
        2386898937694609920 & -- & 0.24 & $-0.506\pm0.074$ & $0.751\pm0.112$ & $40.70\pm5.51$ & 6.89e-05 & 1.69e-06 \\
        6780592417841129472 & K & 0.70 & $-2.243\pm0.045$ & $0.789\pm0.127$ & $40.12\pm0.78$ & 1.92e-04 & 4.78e-06 \\
        4189858726030433792 & M & 0.46 & $-0.608\pm0.006$ & $0.805\pm0.009$ & $67.20\pm0.61$ & 7.23e-05 & 1.08e-06 \\
        4093353529682577152 & K & 0.70 & $-3.943\pm0.203$ & $0.896\pm0.288$ & $60.83\pm3.12$ & 1.10e-04 & 1.81e-06 \\
        6731311271597112576 & M & 0.34 & $-2.387\pm1.015$ & $0.938\pm2.001$ & $36.56\pm9.79$ & 8.48e-05 & 2.32e-06 \\
        6570039342736534784 & G & 1.02 & $-0.837\pm0.003$ & $0.942\pm0.053$ & $49.34\pm0.12$ & 1.89e-04 & 3.83e-06 \\
        5079819487844050432 & G & 0.90 & $-1.811\pm0.020$ & $0.952\pm0.243$ & $78.03\pm0.70$ & 1.04e-04 & 1.33e-06 \\
        6866660813675237248 & K & 0.81 & $-3.255\pm0.110$ & $0.979\pm0.069$ & $72.60\pm2.39$ & 9.79e-05 & 1.35e-06 \\
        5254061535106490112 & M & 0.13 & $-0.030\pm0.000$ & $0.988\pm0.005$ & $157.05\pm1.23$ & 6.97e-06 & 4.44e-08 \\
        6512303781003214464 & K & 0.70 & $-1.035\pm0.004$ & $0.992\pm0.086$ & $71.97\pm0.19$ & 8.41e-05 & 1.17e-06 \\
        6575069780231675264 & M & 0.28 & $-0.225\pm0.002$ & $0.994\pm0.008$ & $32.96\pm0.24$ & 7.30e-05 & 2.21e-06 \\
        5909461003111201792 & M & 0.41 & $-0.779\pm0.027$ & $1.037\pm0.071$ & $111.33\pm3.77$ & 3.05e-05 & 2.74e-07 \\
        6849531247148752512 & K & 0.67 & $-2.066\pm0.053$ & $1.059\pm0.050$ & $76.17\pm1.91$ & 7.14e-05 & 9.37e-07 \\
        6586084515921321728 & M & 0.45 & $-3.076\pm0.173$ & $1.111\pm0.152$ & $38.33\pm2.12$ & 9.01e-05 & 2.35e-06 \\
        6734485286788609664 & M & 0.36 & $-0.082\pm0.000$ & $1.119\pm0.002$ & $164.60\pm0.27$ & 1.70e-05 & 1.03e-07 \\
        2661803855688028800 & M & 0.54 & $-0.253\pm0.001$ & $1.151\pm0.006$ & $63.64\pm0.20$ & 6.28e-05 & 9.87e-07 \\
        6794047652729201024 & M & 0.48 & $-0.196\pm0.001$ & $1.168\pm0.009$ & $48.07\pm0.36$ & 7.42e-05 & 1.54e-06 \\
        4307416074032507136 & M & 0.55 & $-1.705\pm0.053$ & $1.189\pm0.045$ & $50.19\pm1.57$ & 7.91e-05 & 1.58e-06 \\
        6813799318266090496 & M & 0.22 & $-2.214\pm0.415$ & $1.218\pm0.203$ & $33.48\pm5.58$ & 4.70e-05 & 1.40e-06 \\
        6374877128317186048 & K & 0.88 & $-1.350\pm0.005$ & $1.224\pm0.078$ & $87.68\pm0.21$ & 7.02e-05 & 8.00e-07 \\
        4072260704719970944 & K & 0.59 & $-0.220\pm0.001$ & $1.231\pm0.004$ & $78.33\pm0.22$ & 5.30e-05 & 6.77e-07 \\
        4075141768785646848 & M & 0.26 & $-0.053\pm0.001$ & $1.253\pm0.012$ & $49.94\pm0.48$ & 3.50e-05 & 7.02e-07 \\
        5624302662446291072 & K & 0.73 & $-1.544\pm0.020$ & $1.253\pm0.274$ & $124.53\pm1.53$ & 4.02e-05 & 3.23e-07 \\
        6697858840773435776 & -- & 0.26 & $-0.854\pm0.025$ & $1.255\pm0.070$ & $51.01\pm1.48$ & 3.43e-05 & 6.73e-07 \\
        6861605946404195456 & M & 0.45 & $-4.128\pm0.380$ & $1.267\pm0.220$ & $33.70\pm2.95$ & 9.16e-05 & 2.72e-06 \\
        4083781765499334400 & M & 0.52 & $-1.206\pm0.042$ & $1.276\pm0.065$ & $117.40\pm4.08$ & 3.00e-05 & 2.56e-07 \\
        6640420937456214400 & M & 0.23 & $-0.512\pm0.014$ & $1.289\pm0.040$ & $135.15\pm3.79$ & 1.13e-05 & 8.36e-08 \\
        6731311275891314304 & G & 0.98 & $-4.339\pm0.081$ & $1.334\pm0.197$ & $20.08\pm0.37$ & 3.15e-04 & 1.57e-05 \\
        6851353515873281536 & M & 0.36 & $-1.818\pm0.110$ & $1.334\pm0.105$ & $39.94\pm2.36$ & 5.78e-05 & 1.45e-06 \\
        6690652126172090752 & K & 0.72 & $-1.220\pm0.008$ & $1.410\pm0.074$ & $68.06\pm0.38$ & 6.47e-05 & 9.51e-07 \\
        6748488052701979136 & M & 0.44 & $-1.689\pm0.089$ & $1.451\pm0.068$ & $54.13\pm2.82$ & 4.77e-05 & 8.82e-07 \\
        6602563858757140992 & K & 0.25 & $-0.783\pm0.036$ & $1.557\pm0.174$ & $39.25\pm1.81$ & 3.57e-05 & 9.10e-07 \\
        5143621433184275200 & M & 0.55 & $-0.593\pm0.003$ & $1.582\pm0.018$ & $38.78\pm0.20$ & 7.66e-05 & 1.98e-06 \\
        5976042887501445376 & M & 0.40 & $-0.308\pm0.002$ & $1.597\pm0.014$ & $105.83\pm0.80$ & 2.06e-05 & 1.94e-07 \\
        2331575021572597504 & M & 0.35 & $-0.636\pm0.006$ & $1.601\pm0.026$ & $50.98\pm0.45$ & 3.73e-05 & 7.33e-07 \\
        5853498713190525696 & M & 0.12 & $-0.030\pm0.000$ & $1.605\pm0.005$ & $30.76\pm0.01$ & 2.02e-05 & 6.56e-07 \\
        6713373362161524736 & M & 0.27 & $-0.438\pm0.018$ & $1.619\pm0.069$ & $129.32\pm5.37$ & 1.10e-05 & 8.54e-08 \\
        5032483416324388480 & M & 0.34 & $-1.738\pm0.078$ & $1.630\pm0.086$ & $18.72\pm0.82$ & 9.46e-05 & 5.05e-06 \\
        4283408237650849408 & G & 0.95 & $-0.461\pm0.002$ & $1.672\pm0.045$ & $78.94\pm0.13$ & 6.19e-05 & 7.84e-07 \\
        5944732953871170176 & G & 0.90 & $-1.094\pm0.006$ & $1.672\pm0.025$ & $99.27\pm0.40$ & 4.64e-05 & 4.67e-07 \\
        6650117736660187136 & M & 0.22 & $-0.150\pm0.001$ & $1.688\pm0.010$ & $65.80\pm0.39$ & 1.67e-05 & 2.55e-07 \\
        6771477054134260352 & M & 0.32 & $-0.740\pm0.020$ & $1.716\pm0.047$ & $49.72\pm1.34$ & 3.17e-05 & 6.38e-07 \\
        6818312813497758720 & M & 0.38 & $-0.758\pm0.012$ & $1.759\pm0.029$ & $62.68\pm0.99$ & 2.97e-05 & 4.73e-07 \\
        4079684229322231040 & -- & 0.98 & $-0.134\pm0.000$ & $1.777\pm0.004$ & $93.83\pm0.13$ & 5.05e-05 & 5.38e-07 \\
        6718894388002453120 & -- & 0.97 & $-0.143\pm0.000$ & $1.951\pm0.012$ & $115.01\pm0.12$ & 3.73e-05 & 3.24e-07 \\
        6427708902553822592 & M & 0.42 & $-0.181\pm0.001$ & $1.965\pm0.010$ & $68.44\pm0.16$ & 2.67e-05 & 3.91e-07 \\ \bottomrule
    \end{tabular}
\end{table}

\section{Kinematic Classification of 3I/ATLAS}
\label{kinematic}

\begin{figure}[h!]
    \centering
    \includegraphics[width=0.48\textwidth]{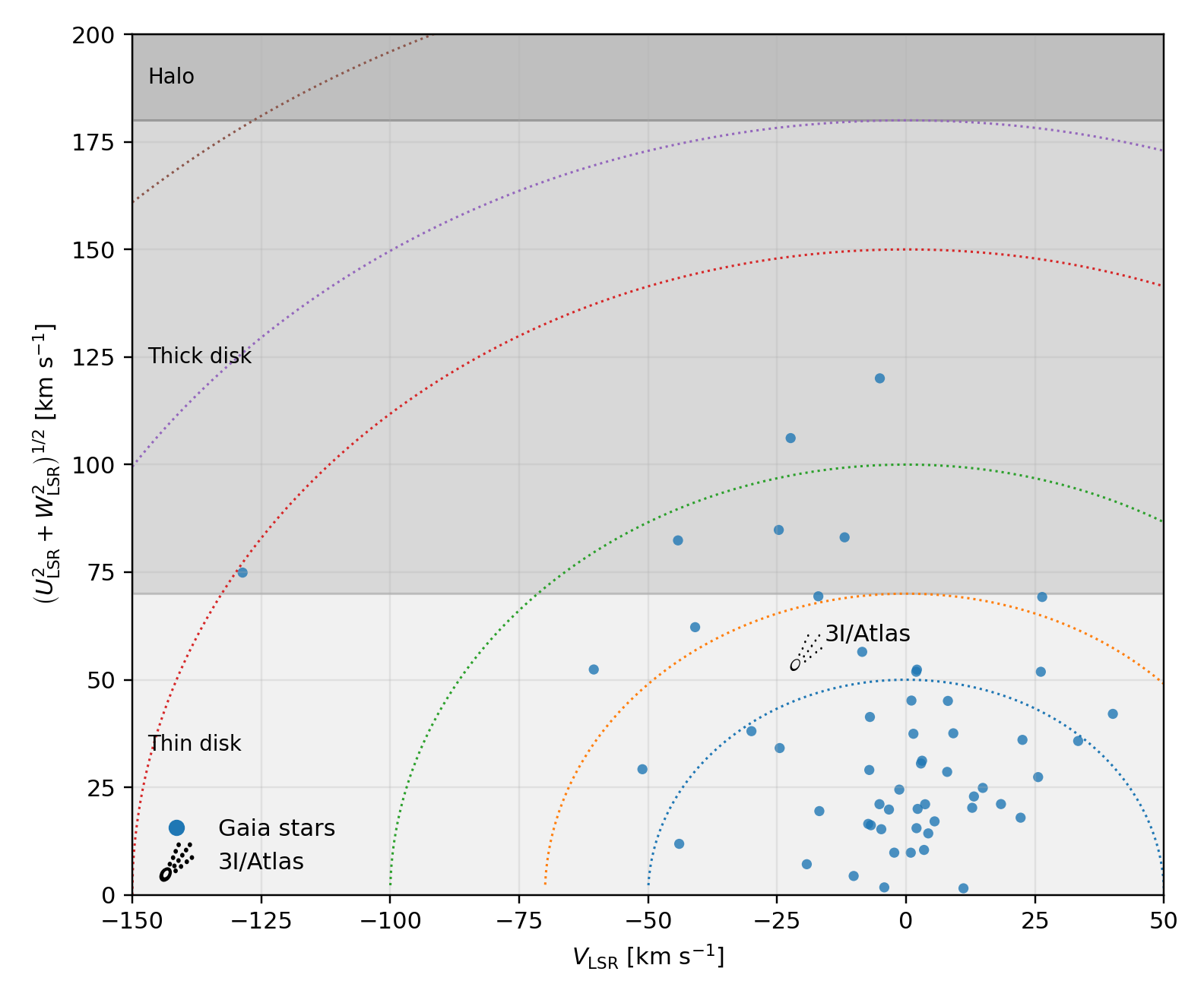}
    \includegraphics[width=0.48\textwidth]{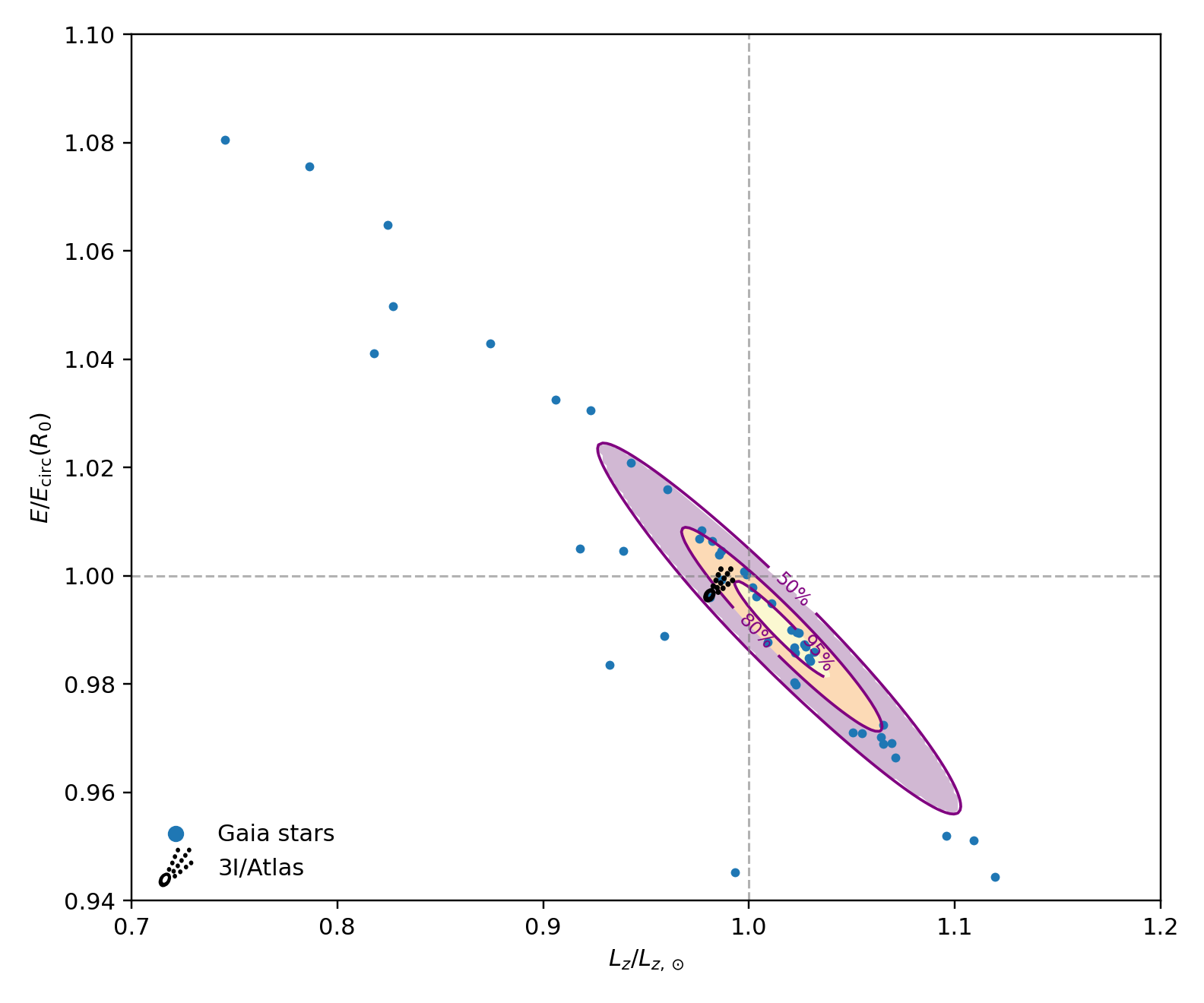}
    \caption{Kinematic diagnostics for 3I/ATLAS. \emph{Left:} Toomre diagram for 3I/ATLAS. The vertical axis shows the quadrature of the radial and vertical velocities ($T$) and the horizontal axis the azimuthal velocity. Shaded regions indicate the thin disk ($T \lesssim 70$ km s$^{-1}$), thick disk ($70 \lesssim T \lesssim 180$ km s$^{-1}$), and halo ($T \gtrsim 180$ km s$^{-1}$). \emph{Right:} Specific orbital energy as function of angular momentum for 3I/ATLAS and nearby \emph{Gaia} stars (Table\ref{tab:encounters}). Background density contours (50, 80, and 95\%) mark the thin-disk locus in the $E$–$L_z$ plane, normalized by the circular orbit at $R_0$. The comet symbol represent 3I/ATLAS while, the blue dots the nearby \emph{Gaia} stars.}
    \label{fig:kinematic}
\end{figure}

The Galactic population membership of 3I/ATLAS can be assessed through three complementary kinematic diagnostics, the Toomre diagram, integrals-of-motion diagram \((E,L_z)\), and the kinematic classification scheme of \citet{Bensby2014}. First, the Toomre diagram compares the vertical velocity  $T=\sqrt{U^2+W^2}$ with the azimuthal velocity $V_{\rm LSR}$ relative to the Local Standard of Rest. In Fig.~\ref{fig:kinematic} left panel, the shaded regions mark the approximate kinematic domains of the thin disk ($T \lesssim 70$ km s$^{-1}$), thick disk ($70 \lesssim T \lesssim 180$ km s$^{-1}$), and halo ($T \gtrsim 180$ km s$^{-1}$), with reference curves of constant peculiar velocity $V_{\rm pec}=\sqrt{U^2+V^2+W^2}$. 3I/ATLAS lies well inside the thin-disk regime, with $T\approx55$ km s$^{-1}$ and $V_{\rm pec}\approx58$ km s$^{-1}$, below the canonical thin/thick-disk boundary at $T\sim70$--$100$ km s$^{-1}$ and with a vertical displacement of $|Z|\sim0.42$ kpc (Fig.~\ref{fig:close_encounters_XZ}). 

Second, we computed the specific orbital energy ($E$) and vertical angular momentum ($L_{z}$) following the methodology in Section ~\ref{sec:methods}. These were evaluated as $E = \tfrac{1}{2}(v_x^2+v_y^2+v_z^2) + \Phi(R,z)$ and $L_z = x\,v_y - y\,v_x$, and normalized by the circular values at the solar radius, $E_{\rm circ}(R_0)$ and $L_{z,\odot}=R_0V_c(R_0)$. In Figure \ref{fig:kinematic} right panel, thin-disk stars cluster near $(L_z/L_{z,\odot}, E/E_{\rm circ}) \approx (1,1)$, while thick-disk and halo stars extend to lower $L_z$ and higher $E$. 3I/ATLAS lies within the thin-disk locus, kinematically indistinguishable from the local disk population.

Finally, we assessed the Galactic population membership of 3I/ATLAS using the kinematic classification scheme of \citet{Bensby2014}, which refines the earlier criteria of \citet{Bensby2003}. For each stellar population $X$ (thin disk, thick disk, halo), the relative likelihood is
\begin{equation}
f_X \;\propto\; X_X\,
\exp\!\left[
-\tfrac{1}{2}\left(
\frac{U^2}{\sigma_{U,X}^2}
+ \frac{(V-V_{\mathrm{asym},X})^2}{\sigma_{V,X}^2}
+ \frac{W^2}{\sigma_{W,X}^2}
\right)\right],
\end{equation}
where $(U,V,W)$ are the Galactic velocity components relative to the LSR, $(\sigma_{U,X},\sigma_{V,X},\sigma_{W,X})$ are the velocity dispersions of population $X$, $V_{\mathrm{asym},X}$ is its asymmetric drift, and $X_X$ is the local fractional density of that population. The classification is based on the ratio $\frac{\rm TD}{\rm D} \;=\; \frac{f_{\rm TD}}{f_{\rm D}},$ which measures how much more likely an object is to belong to the thick disk relative to the thin disk. Using the parameters by \citet{Bensby2014} and the measured Galactic velocities of 3I/ATLAS, $(U,V,W)=(-51.2,-19.5,+18.9)$ km s$^{-1}$, we find $\frac{\rm TD}{\rm D} \;\simeq\; 0.04,$ indicating that 3I/ATLAS is about 20 times more likely to belong to the thin disk than the thick disk. 

Taken together, these three diagnostics—velocity space (Toomre diagram), orbital integrals ($E$--$L_z$), and likelihood ratios (Bensby method)—all converge on the same conclusion, 3I/ATLAS is dynamically consistent  with the Galactic thin disk, and shows no kinematic evidence of thick-disk or halo origin.

\section{Comparison with Previous Studies}
\label{comparison_guo}

The recent study by \citet{guo2025} identified 25 stellar encounters with 3I/ATLAS within 1~pc. In our analysis, we recover only 13 of these events. To understand the origin of this discrepancy, we highlight three key methodological differences between the two studies: (i) the adopted initial conditions, (ii) the treatment of \emph{Gaia} systematics, and (iii) the choice of Galactic potential. 

\citet{guo2025} adopted as initial conditions the ecliptic heliocentric coordinates for J2016.0 (inbound orbit) computed by JPL Horizons\footnote{\url{https://ssd.jpl.nasa.gov/horizons/app.html}}. We instead used  the position and velocities of 3I/ATLAS prior to its entry into the Solar System (unbound orbit) as it was calculated by \citet{delafuenteMarcos2025}. To evaluate the impact of this choice, we re-integrated their sources using both our initial conditions and theirs. We found no significant differences in the resulting outputs (a few tenths of a pc, as expected as the velocities differ by $\approx0.1$\,km\,s$^{-1}$).

We adopted a similar approach to \cite{guo2025} when accounting for systematics in proper motions \citep{Cantat-Gaudin2021} and parallaxes \citep{Lindegren2021} (L21); however, \citet{guo2025} applied an additional parallax bias correction specifically tailored to the Galactic plane \citep[$|b|<20^{\circ}$,][D24]{Ding2024}. We found, however, that most of their encounters have $|b|>20^{\circ}$, in which case only the L21 correction is applied. Moreover, even in the worst-case scenario, the offset between L21 and D24 is only $0.01$ mas. Applying this offset to the L21 corrected parallax of the \citet{guo2025} encounters gives a median error in the relative distance between encounters of just $0.05$ pc. Thus, the D24 parallax correction cannot account for the discrepancies between our results.
        
Finally, in our nominal integrations we reproduce only 13 of the encounters reported by \citet{guo2025}; the rest diverge in distance as time increases. Such discrepancies are expected in long look-back orbit calculations, small differences in astrometry or in the adopted Galactic potential remain negligible over $\lesssim$1--2 Myr but amplify over tens of Myr, shifting closest approaches by several Myr or parsecs. This explains why recent, nearby encounters (13 in our case) are robust and consistent with \citet{guo2025} work, whereas longer look-back events are highly sensitive to both measurement errors and model assumptions. A further systematic arises because \citet{guo2025} integrated with the \citet{Zhou2023} potential (Plummer bulge, razor-thin exponential disks, NFW halo), while our nominal runs used the \texttt{MilkyWayPotential2022} from \texttt{Gala} (spherical nucleus and bulge, a sum of Miyamoto-Nagai disks, and a spherical NFW halo). Although both are smooth and axisymmetric, their disk scale lengths and halo normalizations differ, leading to accumulated systematic offsets as time goes on and explaining much of the mismatch. 

To quantify this, we present in Table~\ref{tab:comparison} the nominal encounter distance of 3I/ATLAS with \emph{Gaia} DR3 6741607618172465152 at around -4.5\,Myr, which \cite{guo2025} find to be 0.85\,pc. We test both the initial conditions of \cite{guo2025} and \cite{delafuenteMarcos2025}, and use four potentials implemented in \texttt{Gala:} \texttt{MilkyWayPotential2022} as in our main simulations, \texttt{MilkyWayPotential} \citep{Bovy2015}, \texttt{LM10Potential} \citep{Law2010}, and \texttt{BovyMWPotential} \citep{Bovy2015}. We find nominal values ranging from 5.94 to 9.91\,pc for both initial conditions  \cite{delafuenteMarcos2025} and \cite{guo2025}. While none of these potentials gives a nominal encounter as close as \cite{guo2025} find, the range over several pc suggests that differences in the potential used may indeed be responsible for our discrepant encounters beyond $\approx1$\,Myr. In contrast, the effect of changing the initial conditions is not so pronounced. In short, the robust encounters trace \emph{Gaia}’s astrometric precision, while the fragile ones expose the sensitivity to Galactic mass models.

\begin{table*}[]
    \centering
    \caption{Encounter distances in pc of 3I/ATLAS with \emph{Gaia} DR3 6741607618172465152, using four different potentials in \texttt{Gala} and two different initial velocities of 3I/ATLAS.}
    \scriptsize
    \begin{tabular}{lcccc}
    \toprule
         Initial conditions & MilkyWayPotential2022 & MilkyWayPotential & LM10Potential & BovyMWPotential \\
         \hline
         \cite{delafuenteMarcos2025} & 6.34 & 9.74 & 5.94 & 9.91 \\
         \cite{guo2025} & 8.92 & 10.16 & 6.34 & 10.29\\
         \bottomrule
    \end{tabular}
    \label{tab:comparison}
\end{table*}


\bibliography{3I_Altas_XPC_etal}{}

@ARTICLE{Abdurrouf2022,
  title     = "The seventeenth data release of the Sloan Digital Sky Surveys:
               Complete release of {MaNGA}, {MaStar}, and {APOGEE-2} data",
  author    = "{Abdurro'uf} and Accetta, Katherine and Aerts, Conny and Silva
               Aguirre, V{\'\i}ctor and Ahumada, Romina and Ajgaonkar, Nikhil
               and Filiz Ak, N and Alam, Shadab and Allende Prieto, Carlos and
               Almeida, Andr{\'e}s and Anders, Friedrich and Anderson, Scott F
               and Andrews, Brett H and Anguiano, Borja and Aquino-Ort{\'\i}z,
               Erik and Arag{\'o}n-Salamanca, Alfonso and Argudo-Fern{\'a}ndez,
               Maria and Ata, Metin and Aubert, Marie and Avila-Reese, Vladimir
               and Badenes, Carles and Barb{\'a}, Rodolfo H and Barger, Kat and
               Barrera-Ballesteros, Jorge K and Beaton, Rachael L and Beers,
               Timothy C and Belfiore, Francesco and Bender, Chad F and
               Bernardi, Mariangela and Bershady, Matthew A and Beutler,
               Florian and Bidin, Christian Moni and Bird, Jonathan C and
               Bizyaev, Dmitry and Blanc, Guillermo A and Blanton, Michael R
               and Boardman, Nicholas Fraser and Bolton, Adam S and Boquien,
               M{\'e}d{\'e}ric and Borissova, Jura and Bovy, Jo and Brandt, W N
               and Brown, Jordan and Brownstein, Joel R and Brusa, Marcella and
               Buchner, Johannes and Bundy, Kevin and Burchett, Joseph N and
               {Martin Bureau} and Burgasser, Adam and Cabang, Tuesday K and
               Campbell, Stephanie and Cappellari, Michele and Carlberg, Joleen
               K and Wanderley, F{\'a}bio Carneiro and Carrera, Ricardo and
               Cash, Jennifer and Chen, Yan-Ping and Chen, Wei-Huai and
               Cherinka, Brian and Chiappini, Cristina and Choi, Peter Doohyun
               and Chojnowski, S Drew and Chung, Haeun and Clerc, Nicolas and
               Cohen, Roger E and Comerford, Julia M and Comparat, Johan and da
               Costa, Luiz and Covey, Kevin and Crane, Jeffrey D and
               Cruz-Gonzalez, Irene and Culhane, Connor and Cunha, Katia and
               Dai, Y Sophia and Damke, Guillermo and Darling, Jeremy and
               Davidson, Jr, James W and Davies, Roger and Dawson, Kyle and De
               Lee, Nathan and Diamond-Stanic, Aleksandar M and Cano-D{\'\i}az,
               Mariana and S{\'a}nchez, Helena Dom{\'\i}nguez and Donor, John
               and Duckworth, Chris and Dwelly, Tom and Eisenstein, Daniel J
               and Elsworth, Yvonne P and Emsellem, Eric and Eracleous, Mike
               and Escoffier, Stephanie and Fan, Xiaohui and Farr, Emily and
               Feng, Shuai and Fern{\'a}ndez-Trincado, Jos{\'e} G and Feuillet,
               Diane and Filipp, Andreas and Fillingham, Sean P and Frinchaboy,
               Peter M and Fromenteau, Sebastien and Galbany, Llu{\'\i}s and
               Garc{\'\i}a, Rafael A and Garc{\'\i}a-Hern{\'a}ndez, D A and Ge,
               Junqiang and Geisler, Doug and Gelfand, Joseph and G{\'e}ron,
               Tobias and Gibson, Benjamin J and Goddy, Julian and
               Godoy-Rivera, Diego and Grabowski, Kathleen and Green, Paul J
               and Greener, Michael and Grier, Catherine J and Griffith, Emily
               and Guo, Hong and Guy, Julien and Hadjara, Massinissa and
               Harding, Paul and Hasselquist, Sten and Hayes, Christian R and
               Hearty, Fred and Hern{\'a}ndez, Jes{\'u}s and Hill, Lewis and
               Hogg, David W and Holtzman, Jon A and Horta, Danny and Hsieh,
               Bau-Ching and Hsu, Chin-Hao and Hsu, Yun-Hsin and Huber, Daniel
               and {Marc Huertas-Company} and Hutchinson, Brian and Hwang, Ho
               Seong and Ibarra-Medel, H{\'e}ctor J and Chitham, Jacob Ider and
               Ilha, Gabriele S and Imig, Julie and Jaekle, Will and
               Jayasinghe, Tharindu and Ji, Xihan and Johnson, Jennifer A and
               Jones, Amy and J{\"o}nsson, Henrik and Katkov, Ivan and
               Khalatyan, Dr Arman and Kinemuchi, Karen and Kisku, Shobhit and
               Knapen, Johan H and Kneib, Jean-Paul and Kollmeier, Juna A and
               Kong, Miranda and Kounkel, Marina and Kreckel, Kathryn and
               Krishnarao, Dhanesh and Lacerna, Ivan and Lane, Richard R and
               Langgin, Rachel and Lavender, Ramon and Law, David R and Lazarz,
               Daniel and Leung, Henry W and Leung, Ho-Hin and Lewis, Hannah M
               and Li, Cheng and Li, Ran and Lian, Jianhui and Liang, Fu-Heng
               and Lin, Lihwai and Lin, Yen-Ting and Lin, Sicheng and Lintott,
               Chris and Long, Dan and Longa-Pe{\~n}a, Pen{\'e}lope and
               L{\'o}pez-Cob{\'a}, Carlos and Lu, Shengdong and Lundgren, Britt
               F and Luo, Yuanze and Mackereth, J Ted and de la Macorra, Axel
               and Mahadevan, Suvrath and Majewski, Steven R and Manchado,
               Arturo and Mandeville, Travis and Maraston, Claudia and
               Margalef-Bentabol, Berta and Masseron, Thomas and Masters, Karen
               L and Mathur, Savita and McDermid, Richard M and Mckay, Myles
               and Merloni, Andrea and Merrifield, Michael and Meszaros,
               Szabolcs and Miglio, Andrea and Di Mille, Francesco and Minniti,
               Dante and Minsley, Rebecca and Monachesi, Antonela and Moon,
               Jeongin and Mosser, Benoit and Mulchaey, John and Muna, Demitri
               and Mu{\~n}oz, Ricardo R and Myers, Adam D and Myers, Natalie
               and Nadathur, Seshadri and Nair, Preethi and Nandra, Kirpal and
               Neumann, Justus and Newman, Jeffrey A and Nidever, David L and
               Nikakhtar, Farnik and Nitschelm, Christian and O'Connell, Julia
               E and Garma-Oehmichen, Luis and Luan Souza de Oliveira, Gabriel
               and Olney, Richard and Oravetz, Daniel and Ortigoza-Urdaneta,
               Mario and Osorio, Yeisson and Otter, Justin and Pace, Zachary J
               and Padilla, Nelson and Pan, Kaike and Pan, Hsi-An and Parikh,
               Taniya and Parker, James and Peirani, Sebastien and Pe{\~n}a
               Ram{\'\i}rez, Karla and Penny, Samantha and Percival, Will J and
               Perez-Fournon, Ismael and Pinsonneault, Marc and Poidevin,
               Fr{\'e}d{\'e}rick and Poovelil, Vijith Jacob and Price-Whelan,
               Adrian M and B{\'a}rbara de Andrade Queiroz, Anna and Raddick, M
               Jordan and Ray, Amy and Rembold, Sandro Barboza and Riddle,
               Nicole and Riffel, Rogemar A and Riffel, Rog{\'e}rio and Rix,
               Hans-Walter and Robin, Annie C and Rodr{\'\i}guez-Puebla, Aldo
               and Roman-Lopes, Alexandre and Rom{\'a}n-Z{\'u}{\~n}iga, Carlos
               and Rose, Benjamin and Ross, Ashley J and Rossi, Graziano and
               Rubin, Kate H R and Salvato, Mara and S{\'a}nchez, Seb{\'a}stian
               F and S{\'a}nchez-Gallego, Jos{\'e} R and Sanderson, Robyn and
               Santana Rojas, Felipe Antonio and Sarceno, Edgar and Sarmiento,
               Regina and Sayres, Conor and Sazonova, Elizaveta and Schaefer,
               Adam L and Schiavon, Ricardo and Schlegel, David J and
               Schneider, Donald P and Schultheis, Mathias and Schwope, Axel
               and Serenelli, Aldo and Serna, Javier and Shao, Zhengyi and
               Shapiro, Griffin and Sharma, Anubhav and Shen, Yue and Shetrone,
               Matthew and Shu, Yiping and Simon, Joshua D and Skrutskie, M F
               and Smethurst, Rebecca and Smith, Verne and Sobeck, Jennifer and
               Spoo, Taylor and Sprague, Dani and Stark, David V and Stassun,
               Keivan G and Steinmetz, Matthias and Stello, Dennis and
               Stone-Martinez, Alexander and Storchi-Bergmann, Thaisa and
               Stringfellow, Guy S and Stutz, Amelia and Su, Yung-Chau and
               Taghizadeh-Popp, Manuchehr and Talbot, Michael S and Tayar,
               Jamie and Telles, Eduardo and Teske, Johanna and Thakar, Ani and
               Theissen, Christopher and Tkachenko, Andrew and Thomas, Daniel
               and Tojeiro, Rita and Hernandez Toledo, Hector and Troup,
               Nicholas W and Trump, Jonathan R and Trussler, James and Turner,
               Jacqueline and Tuttle, Sarah and Unda-Sanzana, Eduardo and
               V{\'a}zquez-Mata, Jos{\'e} Antonio and Valentini, Marica and
               Valenzuela, Octavio and Vargas-Gonz{\'a}lez, Jaime and
               Vargas-Maga{\~n}a, Mariana and Alfaro, Pablo Vera and Villanova,
               Sandro and Vincenzo, Fiorenzo and Wake, David and Warfield, Jack
               T and Washington, Jessica Diane and Weaver, Benjamin Alan and
               Weijmans, Anne-Marie and Weinberg, David H and Weiss, Achim and
               Westfall, Kyle B and Wild, Vivienne and Wilde, Matthew C and
               Wilson, John C and Wilson, Robert F and Wilson, Mikayla and
               Wolf, Julien and Wood-Vasey, W M and Yan, Renbin and Zamora,
               Olga and Zasowski, Gail and Zhang, Kai and Zhao, Cheng and
               Zheng, Zheng and Zheng, Zheng and Zhu, Kai",
  journal   = "Astrophys. J. Suppl. Ser.",
  publisher = "American Astronomical Society",
  volume    =  259,
  number    =  2,
  pages     = "35",
  month     =  apr,
  year      =  2022,
  copyright = "http://creativecommons.org/licenses/by/4.0/"
}

@ARTICLE{2022ApJ...935..167A,
       author = {{Astropy Collaboration} and {Price-Whelan}, Adrian M. and {Lim}, Pey Lian and {Earl}, Nicholas and {Starkman}, Nathaniel and {Bradley}, Larry and {Shupe}, David L. and {Patil}, Aarya A. and {Corrales}, Lia and {Brasseur}, C.~E. and {N{\"o}the}, Maximilian and {Donath}, Axel and {Tollerud}, Erik and {Morris}, Brett M. and {Ginsburg}, Adam and {Vaher}, Eero and {Weaver}, Benjamin A. and {Tocknell}, James and {Jamieson}, William and {van Kerkwijk}, Marten H. and {Robitaille}, Thomas P. and {Merry}, Bruce and {Bachetti}, Matteo and {G{\"u}nther}, H. Moritz and {Aldcroft}, Thomas L. and {Alvarado-Montes}, Jaime A. and {Archibald}, Anne M. and {B{\'o}di}, Attila and {Bapat}, Shreyas and {Barentsen}, Geert and {Baz{\'a}n}, Juanjo and {Biswas}, Manish and {Boquien}, M{\'e}d{\'e}ric and {Burke}, D.~J. and {Cara}, Daria and {Cara}, Mihai and {Conroy}, Kyle E. and {Conseil}, Simon and {Craig}, Matthew W. and {Cross}, Robert M. and {Cruz}, Kelle L. and {D'Eugenio}, Francesco and {Dencheva}, Nadia and {Devillepoix}, Hadrien A.~R. and {Dietrich}, J{\"o}rg P. and {Eigenbrot}, Arthur Davis and {Erben}, Thomas and {Ferreira}, Leonardo and {Foreman-Mackey}, Daniel and {Fox}, Ryan and {Freij}, Nabil and {Garg}, Suyog and {Geda}, Robel and {Glattly}, Lauren and {Gondhalekar}, Yash and {Gordon}, Karl D. and {Grant}, David and {Greenfield}, Perry and {Groener}, Austen M. and {Guest}, Steve and {Gurovich}, Sebastian and {Handberg}, Rasmus and {Hart}, Akeem and {Hatfield-Dodds}, Zac and {Homeier}, Derek and {Hosseinzadeh}, Griffin and {Jenness}, Tim and {Jones}, Craig K. and {Joseph}, Prajwel and {Kalmbach}, J. Bryce and {Karamehmetoglu}, Emir and {Ka{\l}uszy{\'n}ski}, Miko{\l}aj and {Kelley}, Michael S.~P. and {Kern}, Nicholas and {Kerzendorf}, Wolfgang E. and {Koch}, Eric W. and {Kulumani}, Shankar and {Lee}, Antony and {Ly}, Chun and {Ma}, Zhiyuan and {MacBride}, Conor and {Maljaars}, Jakob M. and {Muna}, Demitri and {Murphy}, N.~A. and {Norman}, Henrik and {O'Steen}, Richard and {Oman}, Kyle A. and {Pacifici}, Camilla and {Pascual}, Sergio and {Pascual-Granado}, J. and {Patil}, Rohit R. and {Perren}, Gabriel I. and {Pickering}, Timothy E. and {Rastogi}, Tanuj and {Roulston}, Benjamin R. and {Ryan}, Daniel F. and {Rykoff}, Eli S. and {Sabater}, Jose and {Sakurikar}, Parikshit and {Salgado}, Jes{\'u}s and {Sanghi}, Aniket and {Saunders}, Nicholas and {Savchenko}, Volodymyr and {Schwardt}, Ludwig and {Seifert-Eckert}, Michael and {Shih}, Albert Y. and {Jain}, Anany Shrey and {Shukla}, Gyanendra and {Sick}, Jonathan and {Simpson}, Chris and {Singanamalla}, Sudheesh and {Singer}, Leo P. and {Singhal}, Jaladh and {Sinha}, Manodeep and {Sip{\H{o}}cz}, Brigitta M. and {Spitler}, Lee R. and {Stansby}, David and {Streicher}, Ole and {{\v{S}}umak}, Jani and {Swinbank}, John D. and {Taranu}, Dan S. and {Tewary}, Nikita and {Tremblay}, Grant R. and {de Val-Borro}, Miguel and {Van Kooten}, Samuel J. and {Vasovi{\'c}}, Zlatan and {Verma}, Shresth and {de Miranda Cardoso}, Jos{\'e} Vin{\'\i}cius and {Williams}, Peter K.~G. and {Wilson}, Tom J. and {Winkel}, Benjamin and {Wood-Vasey}, W.~M. and {Xue}, Rui and {Yoachim}, Peter and {Zhang}, Chen and {Zonca}, Andrea and {Astropy Project Contributors}},
        title = "{The Astropy Project: Sustaining and Growing a Community-oriented Open-source Project and the Latest Major Release (v5.0) of the Core Package}",
      journal = {\apj},
         year = 2022,
        month = aug,
       volume = {935},
       number = {2},
          eid = {167},
        pages = {167},
          doi = {10.3847/1538-4357/ac7c74},
archivePrefix = {arXiv},
       eprint = {2206.14220},
 primaryClass = {astro-ph.IM},
       adsurl = {https://ui.adsabs.harvard.edu/abs/2022ApJ...935..167A}
}

@ARTICLE{2018AJ....156..123A,
       author = {{Astropy Collaboration} and {Price-Whelan}, A.~M. and {Sip{\H{o}}cz}, B.~M. and {G{\"u}nther}, H.~M. and {Lim}, P.~L. and {Crawford}, S.~M. and {Conseil}, S. and {Shupe}, D.~L. and {Craig}, M.~W. and {Dencheva}, N. and {Ginsburg}, A. and {VanderPlas}, J.~T. and {Bradley}, L.~D. and {P{\'e}rez-Su{\'a}rez}, D. and {de Val-Borro}, M. and {Aldcroft}, T.~L. and {Cruz}, K.~L. and {Robitaille}, T.~P. and {Tollerud}, E.~J. and {Ardelean}, C. and {Babej}, T. and {Bach}, Y.~P. and {Bachetti}, M. and {Bakanov}, A.~V. and {Bamford}, S.~P. and {Barentsen}, G. and {Barmby}, P. and {Baumbach}, A. and {Berry}, K.~L. and {Biscani}, F. and {Boquien}, M. and {Bostroem}, K.~A. and {Bouma}, L.~G. and {Brammer}, G.~B. and {Bray}, E.~M. and {Breytenbach}, H. and {Buddelmeijer}, H. and {Burke}, D.~J. and {Calderone}, G. and {Cano Rodr{\'\i}guez}, J.~L. and {Cara}, M. and {Cardoso}, J.~V.~M. and {Cheedella}, S. and {Copin}, Y. and {Corrales}, L. and {Crichton}, D. and {D'Avella}, D. and {Deil}, C. and {Depagne}, {\'E}. and {Dietrich}, J.~P. and {Donath}, A. and {Droettboom}, M. and {Earl}, N. and {Erben}, T. and {Fabbro}, S. and {Ferreira}, L.~A. and {Finethy}, T. and {Fox}, R.~T. and {Garrison}, L.~H. and {Gibbons}, S.~L.~J. and {Goldstein}, D.~A. and {Gommers}, R. and {Greco}, J.~P. and {Greenfield}, P. and {Groener}, A.~M. and {Grollier}, F. and {Hagen}, A. and {Hirst}, P. and {Homeier}, D. and {Horton}, A.~J. and {Hosseinzadeh}, G. and {Hu}, L. and {Hunkeler}, J.~S. and {Ivezi{\'c}}, {\v{Z}}. and {Jain}, A. and {Jenness}, T. and {Kanarek}, G. and {Kendrew}, S. and {Kern}, N.~S. and {Kerzendorf}, W.~E. and {Khvalko}, A. and {King}, J. and {Kirkby}, D. and {Kulkarni}, A.~M. and {Kumar}, A. and {Lee}, A. and {Lenz}, D. and {Littlefair}, S.~P. and {Ma}, Z. and {Macleod}, D.~M. and {Mastropietro}, M. and {McCully}, C. and {Montagnac}, S. and {Morris}, B.~M. and {Mueller}, M. and {Mumford}, S.~J. and {Muna}, D. and {Murphy}, N.~A. and {Nelson}, S. and {Nguyen}, G.~H. and {Ninan}, J.~P. and {N{\"o}the}, M. and {Ogaz}, S. and {Oh}, S. and {Parejko}, J.~K. and {Parley}, N. and {Pascual}, S. and {Patil}, R. and {Patil}, A.~A. and {Plunkett}, A.~L. and {Prochaska}, J.~X. and {Rastogi}, T. and {Reddy Janga}, V. and {Sabater}, J. and {Sakurikar}, P. and {Seifert}, M. and {Sherbert}, L.~E. and {Sherwood-Taylor}, H. and {Shih}, A.~Y. and {Sick}, J. and {Silbiger}, M.~T. and {Singanamalla}, S. and {Singer}, L.~P. and {Sladen}, P.~H. and {Sooley}, K.~A. and {Sornarajah}, S. and {Streicher}, O. and {Teuben}, P. and {Thomas}, S.~W. and {Tremblay}, G.~R. and {Turner}, J.~E.~H. and {Terr{\'o}n}, V. and {van Kerkwijk}, M.~H. and {de la Vega}, A. and {Watkins}, L.~L. and {Weaver}, B.~A. and {Whitmore}, J.~B. and {Woillez}, J. and {Zabalza}, V. and {Astropy Contributors}},
        title = "{The Astropy Project: Building an Open-science Project and Status of the v2.0 Core Package}",
      journal = {\aj},
         year = 2018,
        month = sep,
       volume = {156},
       number = {3},
          eid = {123},
        pages = {123},
          doi = {10.3847/1538-3881/aabc4f},
archivePrefix = {arXiv},
       eprint = {1801.02634},
 primaryClass = {astro-ph.IM},
       adsurl = {https://ui.adsabs.harvard.edu/abs/2018AJ....156..123A}
}

@ARTICLE{2013A&A...558A..33A,
       author = {{Astropy Collaboration} and {Robitaille}, Thomas P. and
         {Tollerud}, Erik J. and {Greenfield}, Perry and {Droettboom}, Michael and
         {Bray}, Erik and {Aldcroft}, Tom and {Davis}, Matt and
         {Ginsburg}, Adam and {Price-Whelan}, Adrian M. and
         {Kerzendorf}, Wolfgang E. and {Conley}, Alexander and {Crighton}, Neil and
         {Barbary}, Kyle and {Muna}, Demitri and {Ferguson}, Henry and
         {Grollier}, Fr{\'e}d{\'e}ric and {Parikh}, Madhura M. and
         {Nair}, Prasanth H. and {Unther}, Hans M. and {Deil}, Christoph and
         {Woillez}, Julien and {Conseil}, Simon and {Kramer}, Roban and
         {Turner}, James E.~H. and {Singer}, Leo and {Fox}, Ryan and
         {Weaver}, Benjamin A. and {Zabalza}, Victor and {Edwards}, Zachary I. and
         {Azalee Bostroem}, K. and {Burke}, D.~J. and {Casey}, Andrew R. and
         {Crawford}, Steven M. and {Dencheva}, Nadia and {Ely}, Justin and
         {Jenness}, Tim and {Labrie}, Kathleen and {Lim}, Pey Lian and
         {Pierfederici}, Francesco and {Pontzen}, Andrew and {Ptak}, Andy and
         {Refsdal}, Brian and {Servillat}, Mathieu and {Streicher}, Ole},
        title = "{Astropy: A community Python package for astronomy}",
      journal = {\aap},
         year = "2013",
        month = "Oct",
       volume = {558},
          eid = {A33},
        pages = {A33},
          doi = {10.1051/0004-6361/201322068},
archivePrefix = {arXiv},
       eprint = {1307.6212},
 primaryClass = {astro-ph.IM},
       adsurl = {https://ui.adsabs.harvard.edu/abs/2013A&A...558A..33A}
}

@ARTICLE{Buder2021,
  title     = "The {GALAH+} survey: Third data release",
  author    = "Buder, Sven and Sharma, Sanjib and Kos, Janez and Amarsi, Anish
               M and Nordlander, Thomas and Lind, Karin and Martell, Sarah L
               and Asplund, Martin and Bland-Hawthorn, Joss and Casey, Andrew R
               and De Silva, Gayandhi M and D'Orazi, Valentina and Freeman, Ken
               C and Hayden, Michael R and Lewis, Geraint F and Lin, Jane and
               Schlesinger, Katharine J and Simpson, Jeffrey D and Stello,
               Dennis and Zucker, Daniel B and Zwitter, Toma{\v z} and Beeson,
               Kevin L and Buck, Tobias and Casagrande, Luca and Clark, Jake T
               and {\v C}otar, Klemen and Da Costa, Gary S and de Grijs,
               Richard and Feuillet, Diane and Horner, Jonathan and Kafle,
               Prajwal R and Khanna, Shourya and Kobayashi, Chiaki and Liu, Fan
               and Montet, Benjamin T and Nandakumar, Govind and Nataf, David M
               and Ness, Melissa K and Spina, Lorenzo and Tepper-Garc{\'\i}a,
               Thor and Ting(丁源森), Yuan-Sen and Traven, Gregor and Vogrin{\v
               c}i{\v c}, Rok and Wittenmyer, Robert A and Wyse, Rosemary F G
               and {\v Z}erjal, Maru{\v s}a and {GALAH Collaboration}",
  journal   = "Mon. Not. R. Astron. Soc.",
  publisher = "Oxford University Press (OUP)",
  volume    =  506,
  number    =  1,
  pages     = "150--201",
  month     =  jul,
  year      =  2021,
  copyright = "https://academic.oup.com/journals/pages/open\_access/funder\_policies/chorus/standard\_publication\_model",
  language  = "en"
}

@ARTICLE{delafuenteMarcos2025,
       author = {{de la Fuente Marcos}, R. and {Alarcon}, M.~R. and {Licandro}, J. and {Serra-Ricart}, M. and {de Le{\'o}n}, J. and {de la Fuente Marcos}, C. and {Lombardi}, G. and {Tejero}, A. and {Cabrera-Lavers}, A. and {Guerra Arencibia}, S. and {Ruiz Cejudo}, I.},
        title = "{Assessing interstellar comet 3I/ATLAS with the 10.4 m Gran Telescopio Canarias and the Two-meter Twin Telescope}",
      journal = {\aap},
         year = 2025,
        month = aug,
       volume = {700},
          eid = {L9},
        pages = {L9},
          doi = {10.1051/0004-6361/202556439},
archivePrefix = {arXiv},
       eprint = {2507.12922},
 primaryClass = {astro-ph.EP},
       adsurl = {https://ui.adsabs.harvard.edu/abs/2025A&A...700L...9D}
}

@ARTICLE{GRAVITY2018,
       author = {{GRAVITY Collaboration} and {Abuter}, R. and {Amorim}, A. and {Anugu}, N. and {Baub{\"o}ck}, M. and {Benisty}, M. and {Berger}, J.~P. and {Blind}, N. and {Bonnet}, H. and {Brandner}, W. and {Buron}, A. and {Collin}, C. and {Chapron}, F. and {Cl{\'e}net}, Y. and {Coud{\'e} Du Foresto}, V. and {de Zeeuw}, P.~T. and {Deen}, C. and {Delplancke-Str{\"o}bele}, F. and {Dembet}, R. and {Dexter}, J. and {Duvert}, G. and {Eckart}, A. and {Eisenhauer}, F. and {Finger}, G. and {F{\"o}rster Schreiber}, N.~M. and {F{\'e}dou}, P. and {Garcia}, P. and {Garcia Lopez}, R. and {Gao}, F. and {Gendron}, E. and {Genzel}, R. and {Gillessen}, S. and {Gordo}, P. and {Habibi}, M. and {Haubois}, X. and {Haug}, M. and {Hau{\ss}mann}, F. and {Henning}, Th. and {Hippler}, S. and {Horrobin}, M. and {Hubert}, Z. and {Hubin}, N. and {Jimenez Rosales}, A. and {Jochum}, L. and {Jocou}, K. and {Kaufer}, A. and {Kellner}, S. and {Kendrew}, S. and {Kervella}, P. and {Kok}, Y. and {Kulas}, M. and {Lacour}, S. and {Lapeyr{\`e}re}, V. and {Lazareff}, B. and {Le Bouquin}, J. -B. and {L{\'e}na}, P. and {Lippa}, M. and {Lenzen}, R. and {M{\'e}rand}, A. and {M{\"u}ler}, E. and {Neumann}, U. and {Ott}, T. and {Palanca}, L. and {Paumard}, T. and {Pasquini}, L. and {Perraut}, K. and {Perrin}, G. and {Pfuhl}, O. and {Plewa}, P.~M. and {Rabien}, S. and {Ram{\'\i}rez}, A. and {Ramos}, J. and {Rau}, C. and {Rodr{\'\i}guez-Coira}, G. and {Rohloff}, R. -R. and {Rousset}, G. and {Sanchez-Bermudez}, J. and {Scheithauer}, S. and {Sch{\"o}ller}, M. and {Schuler}, N. and {Spyromilio}, J. and {Straub}, O. and {Straubmeier}, C. and {Sturm}, E. and {Tacconi}, L.~J. and {Tristram}, K.~R.~W. and {Vincent}, F. and {von Fellenberg}, S. and {Wank}, I. and {Waisberg}, I. and {Widmann}, F. and {Wieprecht}, E. and {Wiest}, M. and {Wiezorrek}, E. and {Woillez}, J. and {Yazici}, S. and {Ziegler}, D. and {Zins}, G.},
        title = "{Detection of the gravitational redshift in the orbit of the star S2 near the Galactic centre massive black hole}",
      journal = {\aap},
         year = 2018,
        month = jul,
       volume = {615},
          eid = {L15},
        pages = {L15},
          doi = {10.1051/0004-6361/201833718},
archivePrefix = {arXiv},
       eprint = {1807.09409},
 primaryClass = {astro-ph.GA},
       adsurl = {https://ui.adsabs.harvard.edu/abs/2018A&A...615L..15G}
}

@ARTICLE{Hourihane2023,
  title     = "The {\textit{Gaia}-ESO} Survey: Homogenisation of stellar
               parameters and elemental abundances",
  author    = "Hourihane, A and Fran{\c c}ois, P and Worley, C C and Magrini, L
               and Gonneau, A and Casey, A R and Gilmore, G and Randich, S and
               Sacco, G G and Recio-Blanco, A and Korn, A J and Allende Prieto,
               C and Smiljanic, R and Blomme, R and Bragaglia, A and Walton, N
               A and Van Eck, S and Bensby, T and Lanzafame, A and Frasca, A
               and Franciosini, E and Damiani, F and Lind, K and Bergemann, M
               and Bonifacio, P and Hill, V and Lobel, A and Montes, D and
               Feuillet, D K and Tautvai{\v s}ien{\.e}, G and Guiglion, G and
               Tabernero, H M and Gonz{\'a}lez Hern{\'a}ndez, J I and Gebran, M
               and Van der Swaelmen, M and Mikolaitis, {\v S} and Daflon, S and
               Merle, T and Morel, T and Lewis, J R and Gonz{\'a}lez Solares, E
               A and Murphy, D N A and Jeffries, R D and Jackson, R J and
               Feltzing, S and Prusti, T and Carraro, G and Biazzo, K and
               Prisinzano, L and Jofr{\'e}, P and Zaggia, S and Drazdauskas, A
               and Stonkut{\'e}, E and Marfil, E and Jim{\'e}nez-Esteban, F and
               Mahy, L and Guti{\'e}rrez Albarr{\'a}n, M L and Berlanas, S R
               and Santos, W and Morbidelli, L and Spina, L and Minkevi{\v
               c}i{\=u}t{\.e}, R",
  journal   = "Astron. Astrophys.",
  publisher = "EDP Sciences",
  volume    =  676,
  pages     = "A129",
  month     =  aug,
  year      =  2023,
  copyright = "https://creativecommons.org/licenses/by/4.0"
}

@ARTICLE{Katz2023,
  title     = "\textit{Gaia} Data Release 3",
  author    = "Katz, D and Sartoretti, P and Guerrier, A and Panuzzo, P and
               Seabroke, G M and Th{\'e}venin, F and Cropper, M and Benson, K
               and Blomme, R and Haigron, R and Marchal, O and Smith, M and
               Baker, S and Chemin, L and Damerdji, Y and David, M and Dolding,
               C and Fr{\'e}mat, Y and Gosset, E and Jan{\ss}en, K and
               Jasniewicz, G and Lobel, A and Plum, G and Samaras, N and
               Snaith, O and Soubiran, C and Vanel, O and Zwitter, T and
               Antoja, T and Arenou, F and Babusiaux, C and Brouillet, N and
               Caffau, E and Di Matteo, P and Fabre, C and Fabricius, C and
               Fragkoudi, F and Haywood, M and Huckle, H E and Hottier, C and
               Lasne, Y and Leclerc, N and Mastrobuono-Battisti, A and Royer, F
               and Teyssier, D and Zorec, J and Crifo, F and Jean-Antoine
               Piccolo, A and Turon, C and Viala, Y",
  journal   = "Astron. Astrophys.",
  publisher = "EDP Sciences",
  volume    =  674,
  pages     = "A5",
  month     =  jun,
  year      =  2023,
  copyright = "https://creativecommons.org/licenses/by/4.0"
}

@dataset{Luo2019,
       author = {{Luo}, A. -L. and {Zhao}, Y. -H. and {Zhao}, G. and {et al.}},
        title = "{VizieR Online Data Catalog: LAMOST DR7 catalogs (Luo+, 2019)}",
 howpublished = {VizieR On-line Data Catalog: V/156.  Originally published in: 2019RAA..in.prep..L},
         year = 2022,
        month = mar,
          eid = {V/156},
       adsurl = {https://ui.adsabs.harvard.edu/abs/2022yCat.5156....0L}
}

@ARTICLE{Price2017,
       author = {{Price-Whelan}, Adrian M.},
        title = "{Gala: A Python package for galactic dynamics}",
      journal = {The Journal of Open Source Software},
         year = 2017,
        month = oct,
       volume = {2},
          eid = {388},
        pages = {388},
          doi = {10.21105/joss.00388},
       adsurl = {https://ui.adsabs.harvard.edu/abs/2017JOSS....2..388P}
}

@ARTICLE{Steinmetz2020,
  title     = "The sixth Data Release of the Radial Velocity Experiment (rave).
               {II}. Stellar atmospheric parameters, chemical abundances, and
               distances",
  author    = "Steinmetz, Matthias and Guiglion, Guillaume and McMillan, Paul J
               and Matijevi{\v c}, Gal and Enke, Harry and Kordopatis, Georges
               and Zwitter, Toma{\v z} and Valentini, Marica and Chiappini,
               Cristina and Casagrande, Luca and Wojno, Jennifer and Anguiano,
               Borja and Bienaym{\'e}, Olivier and Bijaoui, Albert and Binney,
               James and Burton, Donna and Cass, Paul and de Laverny, Patrick
               and Fiegert, Kristin and Freeman, Kenneth and Fulbright, Jon P
               and Gibson, Brad K and Gilmore, Gerard and Grebel, Eva K and
               Helmi, Amina and Kunder, Andrea and Munari, Ulisse and Navarro,
               Julio F and Parker, Quentin and Ruchti, Gregory R and
               Recio-Blanco, Alejandra and Reid, Warren and Seabroke, George M
               and Siviero, Alessandro and Siebert, Arnaud and Stupar, Milorad
               and Watson, Fred and Williams, Mary E K and Wyse, Rosemary F G
               and Anders, Friedrich and Antoja, Teresa and Birko, Danijela and
               Bland-Hawthorn, Joss and Bossini, Diego and Garc{\'\i}a, Rafael
               A and Carrillo, Ismael and Chaplin, William J and Elsworth,
               Yvonne and Famaey, Benoit and Gerhard, Ortwin and Jofre, Paula
               and Just, Andreas and Mathur, Savita and Miglio, Andrea and
               Minchev, Ivan and Monari, Giacomo and Mosser, Benoit and Ritter,
               Andreas and Rodrigues, Thaise S and Scholz, Ralf-Dieter and
               Sharma, Sanjib and Sysoliatina, Kseniia and {(The Rave
               collaboration)}",
  journal   = "Astron. J.",
  publisher = "American Astronomical Society",
  volume    =  160,
  number    =  2,
  pages     = "83",
  month     =  aug,
  year      =  2020,
  copyright = "https://iopscience.iop.org/page/copyright"
}

@ARTICLE{Bennett2019,
       author = {{Bennett}, Morgan and {Bovy}, Jo},
        title = "{Vertical waves in the solar neighbourhood in Gaia DR2}",
      journal = {\mnras},
         year = 2019,
        month = jan,
       volume = {482},
       number = {1},
        pages = {1417-1425},
          doi = {10.1093/mnras/sty2813},
archivePrefix = {arXiv},
       eprint = {1809.03507},
 primaryClass = {astro-ph.GA},
       adsurl = {https://ui.adsabs.harvard.edu/abs/2019MNRAS.482.1417B}
}

@ARTICLE{Schonrich2010,
       author = {{Sch{\"o}nrich}, Ralph and {Binney}, James and {Dehnen}, Walter},
        title = "{Local kinematics and the local standard of rest}",
      journal = {\mnras},
         year = 2010,
        month = apr,
       volume = {403},
       number = {4},
        pages = {1829-1833},
          doi = {10.1111/j.1365-2966.2010.16253.x},
archivePrefix = {arXiv},
       eprint = {0912.3693},
 primaryClass = {astro-ph.GA},
       adsurl = {https://ui.adsabs.harvard.edu/abs/2010MNRAS.403.1829S}
}

@ARTICLE{Eilers2019,
       author = {{Eilers}, Anna-Christina and {Hogg}, David W. and {Rix}, Hans-Walter and {Ness}, Melissa K.},
        title = "{The Circular Velocity Curve of the Milky Way from 5 to 25 kpc}",
      journal = {\apj},
         year = 2019,
        month = jan,
       volume = {871},
       number = {1},
          eid = {120},
        pages = {120},
          doi = {10.3847/1538-4357/aaf648},
archivePrefix = {arXiv},
       eprint = {1810.09466},
 primaryClass = {astro-ph.GA},
       adsurl = {https://ui.adsabs.harvard.edu/abs/2019ApJ...871..120E}
}

@ARTICLE{Meech2017,
       author = {{Meech}, Karen J. and {Weryk}, Robert and {Micheli}, Marco and {Kleyna}, Jan T. and {Hainaut}, Olivier R. and {Jedicke}, Robert and {Wainscoat}, Richard J. and {Chambers}, Kenneth C. and {Keane}, Jacqueline V. and {Petric}, Andreea and {Denneau}, Larry and {Magnier}, Eugene and {Berger}, Travis and {Huber}, Mark E. and {Flewelling}, Heather and {Waters}, Chris and {Schunova-Lilly}, Eva and {Chastel}, Serge},
        title = "{A brief visit from a red and extremely elongated interstellar asteroid}",
      journal = {\nat},
         year = 2017,
        month = dec,
       volume = {552},
       number = {7685},
        pages = {378-381},
          doi = {10.1038/nature25020},
       adsurl = {https://ui.adsabs.harvard.edu/abs/2017Natur.552..378M}
}

@ARTICLE{deLeon2019,
       author = {{de Le{\'o}n}, Julia and {Licandro}, Javier and {Serra-Ricart}, Miquel and {Cabrera-Lavers}, Antonio and {Font Serra}, Joan and {Scarpa}, Riccardo and {de la Fuente Marcos}, Carlos and {de la Fuente Marcos}, Ra{\'u}l},
        title = "{Interstellar Visitors: A Physical Characterization of Comet C/2019 Q4 (Borisov) with OSIRIS at the 10.4{\,}m GTC}",
      journal = {Research Notes of the American Astronomical Society},
         year = 2019,
        month = sep,
       volume = {3},
       number = {9},
          eid = {131},
        pages = {131},
          doi = {10.3847/2515-5172/ab449c},
       adsurl = {https://ui.adsabs.harvard.edu/abs/2019RNAAS...3..131D}
}

@ARTICLE{Seligman2025,
       author = {{Seligman}, Darryl Z. and {Micheli}, Marco and {Farnocchia}, Davide and {Denneau}, Larry and {Noonan}, John W. and {Hsieh}, Henry H. and {Santana-Ros}, Toni and {Tonry}, John and {Auchettl}, Katie and {Conversi}, Luca and {Devog{\`e}le}, Maxime and {Faggioli}, Laura and {Feinstein}, Adina D. and {Fenucci}, Marco and {Ferrais}, Marin and {Frincke}, Tessa and {Gillon}, Michael and {Hainaut}, Olivier R. and {Hart}, Kyle and {Hoffman}, Andrew and {Holt}, Carrie E. and {Hoogendam}, Willem B. and {Huber}, Mark E. and {Jehin}, Emmanuel and {Kareta}, Theodore and {Keane}, Jacqueline V. and {Kelley}, Michael S.~P. and {Lister}, Tim and {Mandt}, Kathleen and {Manfroid}, Jean and {Mar{\v{c}}eta}, Du{\v{s}}an and {Meech}, Karen J. and {Amine Miftah}, Mohamed and {Morgan}, Marvin and {Oca{\~n}a}, Francisco and {Pe{\~n}a-Asensio}, Eloy and {Shappee}, Benjamin J. and {Siverd}, Robert J. and {Taylor}, Aster G. and {Tucker}, Michael A. and {Wainscoat}, Richard and {Weryk}, Robert and {Wray}, James J. and {Yaginuma}, Atsuhiro and {Yang}, Bin and {Ye}, Quanzhi and {Zhang}, Qicheng},
        title = "{Discovery and Preliminary Characterization of a Third Interstellar Object: 3I/ATLAS}",
      journal = {\apjl},
         year = 2025,
        month = aug,
       volume = {989},
       number = {2},
          eid = {L36},
        pages = {L36},
          doi = {10.3847/2041-8213/adf49a},
archivePrefix = {arXiv},
       eprint = {2507.02757},
 primaryClass = {astro-ph.EP},
       adsurl = {https://ui.adsabs.harvard.edu/abs/2025ApJ...989L..36S}
}

@ARTICLE{Xing2025,
       author = {{Xing}, Zexi and {Oset}, Shawn and {Noonan}, John and {Bodewits}, Dennis},
        title = "{Water Detection in the Interstellar Object 3I/ATLAS}",
      journal = {arXiv e-prints},
         year = 2025,
        month = aug,
          eid = {arXiv:2508.04675},
        pages = {arXiv:2508.04675},
          doi = {10.48550/arXiv.2508.04675},
archivePrefix = {arXiv},
       eprint = {2508.04675},
 primaryClass = {astro-ph.EP},
       adsurl = {https://ui.adsabs.harvard.edu/abs/2025arXiv250804675X}
}

@ARTICLE{Lisse2025,
      title={SPHEREx Discovery of Strong Water Ice Absorption and an Extended Carbon Dioxide Coma in 3I/ATLAS}, 
      author={C. M. Lisse and Y. P. Bach and S. Bryan and B. P. Crill and A. Cukierman and O. Doré and B. Fabinsky and A. Faisst and P. M. Korngut and G. Melnick and Z. Rustamkulov and V. Tolls and M. Werner and M. L. Sitko and C. Champagne and M. Connelley and J. P. Emery and B. Yang and the SPHEREx Science Team},
      year={2025},
      eprint={2508.15469},
      archivePrefix={arXiv},
      primaryClass={astro-ph.EP},
      url={https://arxiv.org/abs/2508.15469}, 
}

@ARTICLE{Taylor2025,
       author = {{Taylor}, Aster G. and {Seligman}, Darryl Z.},
        title = "{The Kinematic Age of 3I/ATLAS and its Implications for Early Planet Formation}",
      journal = {arXiv e-prints},
         year = 2025,
        month = jul,
          eid = {arXiv:2507.08111},
        pages = {arXiv:2507.08111},
          doi = {10.48550/arXiv.2507.08111},
archivePrefix = {arXiv},
       eprint = {2507.08111},
 primaryClass = {astro-ph.EP},
       adsurl = {https://ui.adsabs.harvard.edu/abs/2025arXiv250708111T}
}

@ARTICLE{kakharov2025,
      title={Galactic Trajectories of Interstellar Objects 1I/'Oumuamua, 2I/Borisov, and 3I/Atlas}, 
      author={Shokhruz Kakharov and Abraham Loeb},
      year={2025},
      eprint={2408.02739},
      archivePrefix={arXiv},
      primaryClass={astro-ph.GA},
      url={https://arxiv.org/abs/2408.02739}, 
}

@ARTICLE{Cantat-Gaudin2021,
       author = {{Cantat-Gaudin}, Tristan and {Brandt}, Timothy D.},
        title = "{Characterizing and correcting the proper motion bias of the bright Gaia EDR3 sources}",
      journal = {\aap},
         year = 2021,
        month = may,
       volume = {649},
          eid = {A124},
        pages = {A124},
          doi = {10.1051/0004-6361/202140807},
archivePrefix = {arXiv},
       eprint = {2103.07432},
 primaryClass = {astro-ph.GA},
       adsurl = {https://ui.adsabs.harvard.edu/abs/2021A&A...649A.124C}
}

@ARTICLE{Lindegren2021,
       author = {{Lindegren}, L. and {Bastian}, U. and {Biermann}, M. and {Bombrun}, A. and {de Torres}, A. and {Gerlach}, E. and {Geyer}, R. and {Hern{\'a}ndez}, J. and {Hilger}, T. and {Hobbs}, D. and {Klioner}, S.~A. and {Lammers}, U. and {McMillan}, P.~J. and {Ramos-Lerate}, M. and {Steidelm{\"u}ller}, H. and {Stephenson}, C.~A. and {van Leeuwen}, F.},
        title = "{Gaia Early Data Release 3. Parallax bias versus magnitude, colour, and position}",
      journal = {\aap},
         year = 2021,
        month = may,
       volume = {649},
          eid = {A4},
        pages = {A4},
          doi = {10.1051/0004-6361/202039653},
archivePrefix = {arXiv},
       eprint = {2012.01742},
 primaryClass = {astro-ph.IM},
       adsurl = {https://ui.adsabs.harvard.edu/abs/2021A&A...649A...4L}
}

@ARTICLE{MaizApellaniz2022,
       author = {{Ma{\'\i}z Apell{\'a}niz}, J.},
        title = "{An estimation of the Gaia EDR3 parallax bias from stellar clusters and Magellanic Clouds data}",
      journal = {\aap},
         year = 2022,
        month = jan,
       volume = {657},
          eid = {A130},
        pages = {A130},
          doi = {10.1051/0004-6361/202142365},
archivePrefix = {arXiv},
       eprint = {2110.01475},
 primaryClass = {astro-ph.IM},
       adsurl = {https://ui.adsabs.harvard.edu/abs/2022A&A...657A.130M}
}

@ARTICLE{PecautMamajek2013,
       author = {{Pecaut}, Mark J. and {Mamajek}, Eric E.},
        title = "{Intrinsic Colors, Temperatures, and Bolometric Corrections of Pre-main-sequence Stars}",
      journal = {\apjs},
         year = 2013,
        month = sep,
       volume = {208},
       number = {1},
          eid = {9},
        pages = {9},
          doi = {10.1088/0067-0049/208/1/9},
archivePrefix = {arXiv},
       eprint = {1307.2657},
 primaryClass = {astro-ph.SR},
       adsurl = {https://ui.adsabs.harvard.edu/abs/2013ApJS..208....9P}
}

@book{Galactic_dynamics08,
   author = {J Binney and S Tremaine},
   journal = {Galactic Dynamics: Second Edition, by James Binney and Scott Tremaine.~ISBN 978-0-691-13026-2 (HB).~Published by Princeton University Press, Princeton, NJ USA, 2008.},
   publisher = {Princeton University Press},
   title = {Galactic Dynamics: Second Edition},
   year = {2008},
}

@ARTICLE{Rickman1976,
       author = {{Rickman}, H.},
        title = "{Stellar Perturbations of Orbits of Long-period Comets and their Significance for Cometary Capture}",
      journal = {Bulletin of the Astronomical Institutes of Czechoslovakia},
         year = 1976,
        month = jan,
       volume = {27},
        pages = {92},
       adsurl = {https://ui.adsabs.harvard.edu/abs/1976BAICz..27...92R}
}

@ARTICLE{PZ_Torres2018,
       author = {{Portegies Zwart}, Simon and {Torres}, Santiago and {Pelupessy}, Inti and {B{\'e}dorf}, Jeroen and {Cai}, Maxwell X.},
        title = "{The origin of interstellar asteroidal objects like 1I/2017 U1 `Oumuamua}",
      journal = {\mnras},
         year = 2018,
        month = sep,
       volume = {479},
       number = {1},
        pages = {L17-L22},
          doi = {10.1093/mnrasl/sly088},
archivePrefix = {arXiv},
       eprint = {1711.03558},
 primaryClass = {astro-ph.EP},
       adsurl = {https://ui.adsabs.harvard.edu/abs/2018MNRAS.479L..17P}
}

@article{Torres2023,
       author = {{{Torres,} S.} and {Naoz}, Smadar and {Li}, Gongjie and {Rose}, Sanaea C.},
        title = "{Raining rocks: an analytical formulation for collision time-scales in planetary systems}",
      journal = {\mnras},
         year = 2023,
        month = sep,
       volume = {524},
       number = {1},
        pages = {1025-1030},
          doi = {10.1093/mnras/stad1923},
archivePrefix = {arXiv},
       eprint = {2110.02269},
 primaryClass = {astro-ph.EP},
       adsurl = {https://ui.adsabs.harvard.edu/abs/2023MNRAS.524.1025T}
}

@ARTICLE{Do2018,
       author = {{Do}, Aaron and {Tucker}, Michael A. and {Tonry}, John},
        title = "{Interstellar Interlopers: Number Density and Origin of {\textquoteleft}Oumuamua-like Objects}",
      journal = {\apjl},
         year = 2018,
        month = mar,
       volume = {855},
       number = {1},
          eid = {L10},
        pages = {L10},
          doi = {10.3847/2041-8213/aaae67},
archivePrefix = {arXiv},
       eprint = {1801.02821},
 primaryClass = {astro-ph.EP},
       adsurl = {https://ui.adsabs.harvard.edu/abs/2018ApJ...855L..10D}
}

@ARTICLE{Brasser2013,
       author = {{Brasser}, R. and {Morbidelli}, A.},
        title = "{Oort cloud and Scattered Disc formation during a late dynamical instability in the Solar System}",
      journal = {\icarus},
         year = 2013,
        month = jul,
       volume = {225},
       number = {1},
        pages = {40-49},
          doi = {10.1016/j.icarus.2013.03.012},
archivePrefix = {arXiv},
       eprint = {1303.3098},
 primaryClass = {astro-ph.EP},
       adsurl = {https://ui.adsabs.harvard.edu/abs/2013Icar..225...40B}
}

@ARTICLE{Raymond2018,
       author = {{Raymond}, Sean N. and {Armitage}, Philip J. and {Veras}, Dimitri and {Quintana}, Elisa V. and {Barclay}, Thomas},
        title = "{Implications of the interstellar object 1I/'Oumuamua for planetary dynamics and planetesimal formation}",
      journal = {\mnras},
         year = 2018,
        month = may,
       volume = {476},
       number = {3},
        pages = {3031-3038},
          doi = {10.1093/mnras/sty468},
archivePrefix = {arXiv},
       eprint = {1711.09599},
 primaryClass = {astro-ph.EP},
       adsurl = {https://ui.adsabs.harvard.edu/abs/2018MNRAS.476.3031R}
}

@ARTICLE{JimenezTorres2011,
       author = {{Jim{\'e}nez-Torres}, Juan J. and {Pichardo}, Barbara and {Lake}, George and {Throop}, Henry},
        title = "{Effect of different stellar galactic environments on planetary discs - I. The solar neighbourhood and the birth cloud of the Sun}",
      journal = {\mnras},
         year = 2011,
        month = dec,
       volume = {418},
       number = {2},
        pages = {1272-1284},
          doi = {10.1111/j.1365-2966.2011.19579.x},
archivePrefix = {arXiv},
       eprint = {1108.2412},
 primaryClass = {astro-ph.EP},
       adsurl = {https://ui.adsabs.harvard.edu/abs/2011MNRAS.418.1272J}
}

@ARTICLE{Higuchi2015,
       author = {{Higuchi}, A. and {Kokubo}, E.},
        title = "{Effect of Stellar Encounters on Comet Cloud Formation}",
      journal = {\aj},
         year = 2015,
        month = jul,
       volume = {150},
       number = {1},
          eid = {26},
        pages = {26},
          doi = {10.1088/0004-6256/150/1/26},
archivePrefix = {arXiv},
       eprint = {1507.00502},
 primaryClass = {astro-ph.EP},
       adsurl = {https://ui.adsabs.harvard.edu/abs/2015AJ....150...26H}
}

@INPROCEEDINGS{Levine2021,
       author = {{Levine}, W.~G. and {Laughlin}, G.~P.},
        title = "{Assessing the Hydrogen Ice Hypothesis for 1I/2017 ('Oumuamua)}",
    booktitle = {American Astronomical Society Meeting Abstracts \#238},
         year = 2021,
       series = {American Astronomical Society Meeting Abstracts},
       volume = {238},
        month = jun,
          eid = {232.07},
        pages = {232.07},
       adsurl = {https://ui.adsabs.harvard.edu/abs/2021AAS...23823207L}
}

@ARTICLE{Rafikov2018,
       author = {{Rafikov}, Roman R.},
        title = "{1I/2017 {\textquoteright}Oumuamua-like Interstellar Asteroids as Possible Messengers from Dead Stars}",
      journal = {\apj},
         year = 2018,
        month = jul,
       volume = {861},
       number = {1},
          eid = {35},
        pages = {35},
          doi = {10.3847/1538-4357/aac5ef},
archivePrefix = {arXiv},
       eprint = {1801.02658},
 primaryClass = {astro-ph.EP},
       adsurl = {https://ui.adsabs.harvard.edu/abs/2018ApJ...861...35R}
}

@ARTICLE{ZhangLin2020,
       author = {{Zhang}, Yun and {Lin}, Douglas N.~C.},
        title = "{Tidal fragmentation as the origin of 1I/2017 U1 (`Oumuamua)}",
      journal = {Nature Astronomy},
         year = 2020,
        month = apr,
       volume = {4},
        pages = {852-860},
          doi = {10.1038/s41550-020-1065-8},
archivePrefix = {arXiv},
       eprint = {2004.07218},
 primaryClass = {astro-ph.EP},
       adsurl = {https://ui.adsabs.harvard.edu/abs/2020NatAs...4..852Z}
}

@ARTICLE{Zhang2018,
       author = {{Zhang}, Qicheng},
        title = "{Prospects for Backtracing 1I/{\textquoteleft}Oumuamua and Future Interstellar Objects}",
      journal = {\apjl},
         year = 2018,
        month = jan,
       volume = {852},
       number = {1},
          eid = {L13},
        pages = {L13},
          doi = {10.3847/2041-8213/aaa2f7},
archivePrefix = {arXiv},
       eprint = {1712.08059},
 primaryClass = {astro-ph.EP},
       adsurl = {https://ui.adsabs.harvard.edu/abs/2018ApJ...852L..13Z}
}

@ARTICLE{Heisler1986,
       author = {{Heisler}, J. and {Tremaine}, S.},
        title = "{The influence of the Galactic tidal field on the Oort comet cloud}",
      journal = {\icarus},
         year = 1986,
        month = jan,
       volume = {65},
       number = {1},
        pages = {13-26},
          doi = {10.1016/0019-1035(86)90060-6},
       adsurl = {https://ui.adsabs.harvard.edu/abs/1986Icar...65...13H}
}

@ARTICLE{Rickman2014,
       author = {{Rickman}, Hans},
        title = "{The Oort Cloud and long-period comets}",
      journal = {\maps},
         year = 2014,
        month = jan,
       volume = {49},
       number = {1},
        pages = {8-20},
          doi = {10.1111/maps.12080},
       adsurl = {https://ui.adsabs.harvard.edu/abs/2014M&PS...49....8R}
}

@ARTICLE{Feng2015,
       author = {{Feng}, Fabo and {Bailer-Jones}, C.~A.~L.},
        title = "{Finding the imprints of stellar encounters in long-period comets}",
      journal = {\mnras},
         year = 2015,
        month = dec,
       volume = {454},
       number = {3},
        pages = {3267-3276},
          doi = {10.1093/mnras/stv2222},
archivePrefix = {arXiv},
       eprint = {1509.07222},
 primaryClass = {astro-ph.EP},
       adsurl = {https://ui.adsabs.harvard.edu/abs/2015MNRAS.454.3267F}
}

@ARTICLE{2016A&A...595A...1G,
       author = {{Gaia Collaboration} and {Prusti}, T. and {de Bruijne}, J.~H.~J. and {Brown}, A.~G.~A. and {Vallenari}, A. and {Babusiaux}, C. and {Bailer-Jones}, C.~A.~L. and {Bastian}, U. and {Biermann}, M. and {Evans}, D.~W. and {Eyer}, L. and {Jansen}, F. and {Jordi}, C. and {Klioner}, S.~A. and {Lammers}, U. and {Lindegren}, L. and {Luri}, X. and {Mignard}, F. and {Milligan}, D.~J. and {Panem}, C. and {Poinsignon}, V. and {Pourbaix}, D. and {Randich}, S. and {Sarri}, G. and {Sartoretti}, P. and {Siddiqui}, H.~I. and {Soubiran}, C. and {Valette}, V. and {van Leeuwen}, F. and {Walton}, N.~A. and {Aerts}, C. and {Arenou}, F. and {Cropper}, M. and {Drimmel}, R. and {H{\o}g}, E. and {Katz}, D. and {Lattanzi}, M.~G. and {O'Mullane}, W. and {Grebel}, E.~K. and {Holland}, A.~D. and {Huc}, C. and {Passot}, X. and {Bramante}, L. and {Cacciari}, C. and {Casta{\~n}eda}, J. and {Chaoul}, L. and {Cheek}, N. and {De Angeli}, F. and {Fabricius}, C. and {Guerra}, R. and {Hern{\'a}ndez}, J. and {Jean-Antoine-Piccolo}, A. and {Masana}, E. and {Messineo}, R. and {Mowlavi}, N. and {Nienartowicz}, K. and {Ord{\'o}{\~n}ez-Blanco}, D. and {Panuzzo}, P. and {Portell}, J. and {Richards}, P.~J. and {Riello}, M. and {Seabroke}, G.~M. and {Tanga}, P. and {Th{\'e}venin}, F. and {Torra}, J. and {Els}, S.~G. and {Gracia-Abril}, G. and {Comoretto}, G. and {Garcia-Reinaldos}, M. and {Lock}, T. and {Mercier}, E. and {Altmann}, M. and {Andrae}, R. and {Astraatmadja}, T.~L. and {Bellas-Velidis}, I. and {Benson}, K. and {Berthier}, J. and {Blomme}, R. and {Busso}, G. and {Carry}, B. and {Cellino}, A. and {Clementini}, G. and {Cowell}, S. and {Creevey}, O. and {Cuypers}, J. and {Davidson}, M. and {De Ridder}, J. and {de Torres}, A. and {Delchambre}, L. and {Dell'Oro}, A. and {Ducourant}, C. and {Fr{\'e}mat}, Y. and {Garc{\'\i}a-Torres}, M. and {Gosset}, E. and {Halbwachs}, J. -L. and {Hambly}, N.~C. and {Harrison}, D.~L. and {Hauser}, M. and {Hestroffer}, D. and {Hodgkin}, S.~T. and {Huckle}, H.~E. and {Hutton}, A. and {Jasniewicz}, G. and {Jordan}, S. and {Kontizas}, M. and {Korn}, A.~J. and {Lanzafame}, A.~C. and {Manteiga}, M. and {Moitinho}, A. and {Muinonen}, K. and {Osinde}, J. and {Pancino}, E. and {Pauwels}, T. and {Petit}, J. -M. and {Recio-Blanco}, A. and {Robin}, A.~C. and {Sarro}, L.~M. and {Siopis}, C. and {Smith}, M. and {Smith}, K.~W. and {Sozzetti}, A. and {Thuillot}, W. and {van Reeven}, W. and {Viala}, Y. and {Abbas}, U. and {Abreu Aramburu}, A. and {Accart}, S. and {Aguado}, J.~J. and {Allan}, P.~M. and {Allasia}, W. and {Altavilla}, G. and {{\'A}lvarez}, M.~A. and {Alves}, J. and {Anderson}, R.~I. and {Andrei}, A.~H. and {Anglada Varela}, E. and {Antiche}, E. and {Antoja}, T. and {Ant{\'o}n}, S. and {Arcay}, B. and {Atzei}, A. and {Ayache}, L. and {Bach}, N. and {Baker}, S.~G. and {Balaguer-N{\'u}{\~n}ez}, L. and {Barache}, C. and {Barata}, C. and {Barbier}, A. and {Barblan}, F. and {Baroni}, M. and {Barrado y Navascu{\'e}s}, D. and {Barros}, M. and {Barstow}, M.~A. and {Becciani}, U. and {Bellazzini}, M. and {Bellei}, G. and {Bello Garc{\'\i}a}, A. and {Belokurov}, V. and {Bendjoya}, P. and {Berihuete}, A. and {Bianchi}, L. and {Bienaym{\'e}}, O. and {Billebaud}, F. and {Blagorodnova}, N. and {Blanco-Cuaresma}, S. and {Boch}, T. and {Bombrun}, A. and {Borrachero}, R. and {Bouquillon}, S. and {Bourda}, G. and {Bouy}, H. and {Bragaglia}, A. and {Breddels}, M.~A. and {Brouillet}, N. and {Br{\"u}semeister}, T. and {Bucciarelli}, B. and {Budnik}, F. and {Burgess}, P. and {Burgon}, R. and {Burlacu}, A. and {Busonero}, D. and {Buzzi}, R. and {Caffau}, E. and {Cambras}, J. and {Campbell}, H. and {Cancelliere}, R. and {Cantat-Gaudin}, T. and {Carlucci}, T. and {Carrasco}, J.~M. and {Castellani}, M. and {Charlot}, P. and {Charnas}, J. and {Charvet}, P. and {Chassat}, F. and {Chiavassa}, A. and {Clotet}, M. and {Cocozza}, G. and {Collins}, R.~S. and {Collins}, P. and {Costigan}, G.},
        title = "{The Gaia mission}",
      journal = {\aap},
         year = 2016,
        month = nov,
       volume = {595},
          eid = {A1},
        pages = {A1},
          doi = {10.1051/0004-6361/201629272},
archivePrefix = {arXiv},
       eprint = {1609.04153},
 primaryClass = {astro-ph.IM},
       adsurl = {https://ui.adsabs.harvard.edu/abs/2016A&A...595A...1G}
}
\bibliographystyle{aasjournalv7}



\end{document}